# Three-Dimensional Damage Characterisation in Dual Phase Steel using Deep Learning

Setareh Medghalchi(1)*,  Ehsan Karimi(1), Sang-Hyeok Lee(1), Benjamin Berkels(2), Ulrich Kerzel(3), Sandra Korte-Kerzel (1)

1. Institute for Physical Metallurgy and Materials Physics, RWTH Aachen University, Aachen, Germany.

2. Institute for Advanced Study in Computational Engineering Science, RWTH Aachen University, Aachen Germany.

3. Data Science and Artificial Intelligence in Materials and Geoscience, Fakultät für Georessourcen und Materialtechnik, RWTH Aachen University, Aachen, Germany.

*e-mail: medghalchi@imm.rwth-aachen.de

## Abstract

High performance sheet metals with a multi-phase microstructure suffer from deformation induced damage formation during forming in the constituent phases but importantly also where these intersect. To capture damage in terms of the physical processes in three dimensions (3D) and its stochastic nature during deformation, two challenges remain to be tackled: First, bridging high resolution analysis towards large scales to consider statistical data and, second, characterising in 3D with a resolution appropriate for sub-micron sized voids at a large scale. Here, we present how this can be achieved using panoramic scanning electron microscopy (SEM), metallographic serial sectioning, and deep-learning assisted automatic image analysis. This brings together the 3D evolution of active damage mechanisms with volumetric and environmental information for thousands of individual damage sites. We also assess potential surface preparation artefacts in 2D analyses. Overall, we find that for the material considered here, a dual phase (DP800) steel, martensite cracking is the dominant but not sole origin of deformation induced damage and that for a quantitative comparison of damage density, metallographic preparation can induce additional surface damage density far exceeding what is commonly induced between uniaxial straining steps.

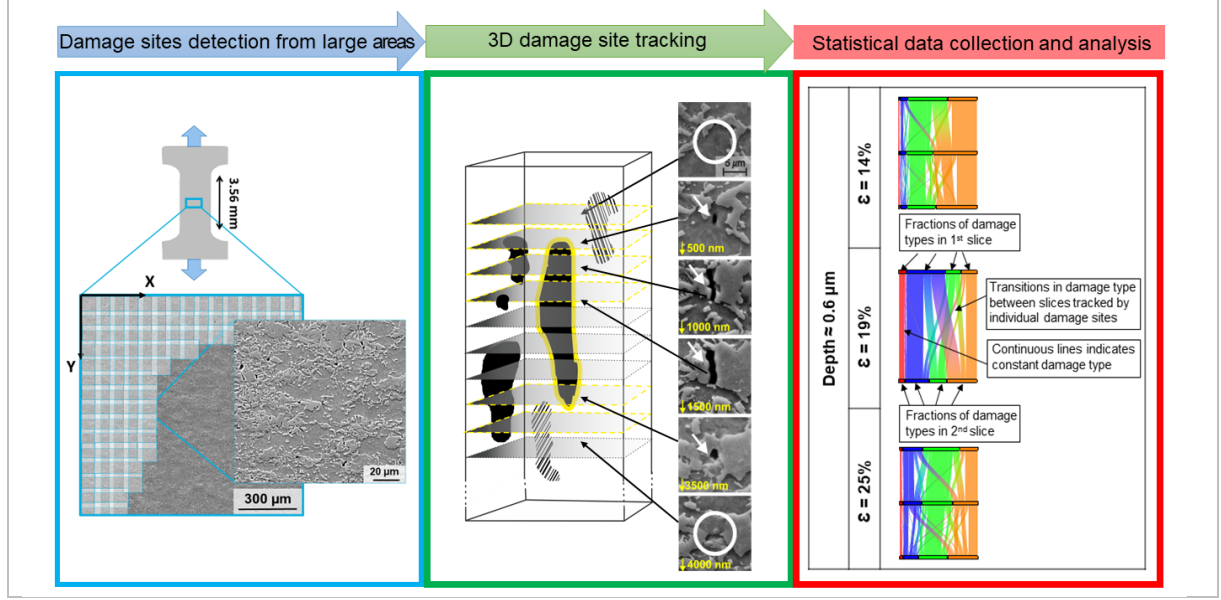

Graphical abstract

# Introduction

During the last few decades, substantial efforts have been made to understand the fundamentals of and model deformation-induced damage and fracture in metallic structures [1-3]. The ultimate aim of this research in terms of application is commonly to improve fuel efficiency in transportation by lightweight design and to increase the safety of crucial components, especially in the automotive industry. Dual phase (DP) steels are among the most applied advanced high strength steels (AHSS) today [4]. They possess a heterogeneous microstructure composed of two phases: hard martensite islands in a relatively softer and more ductile ferrite matrix. This dual microstructure, which can be controlled over a large range of properties based on composition and thermo-mechanical treatment during sheet metal production, gives the DP steels the desired property combination of high strength and formability and accordingly a large energy absorption capacity [5-8] However, the pronounced heterogeneity and mechanical contrast between the two phases also results in their proneness for microscopic damage formation during deformation of the material when the internal co-deformation of the two phases is not compatible [9]. A certain degree of ductile damage inevitably occurs during forming [10] and the formability of dual phase steels can therefore also be controlled by damage induced softening [4].

The design of industrial components will continue to move towards an increasingly integrated and detailed approach in our quest to save energy during production and application and increase safety and profits. Locally variable material properties are one route in which this may be supported. For this, the ability to model the locally remaining strength and strain to failure after sheet forming may be explored and safety margins as well as lightweight design tailored in greater detail. To allow this, improved damage models are essential, as are physical insights to drive the purposeful design of damage tolerant microstructures and dedicated strain paths during processing [10].

One shortcoming of conventional characterisation methods, which ultimately inform such models and design efforts, is that at the scale of an industrially formed sheet, heterogeneity spans far beyond the local co-deformation of the two phases to the length of martensite bands and strain (path) gradients. With this in mind, we previously developed a method to image and analyse the observed damage sites and their underlying nucleation mechanisms across much larger areas at high resolution [11]. We also transferred the underlying convolutional neural networks to the same microstructure deformed under biaxial [12] instead of uniaxial strain and used the damage site statistics in terms of size and number density to investigate the effects of strain and triaxiality on nucleation, growth and coalescence of damage sites [13].

These studies were based on 2D imaging of the sheet material as commonly used in the literature by means of in plane [14] and cross-sectional imaging [3] applied either post-mortem or in-situ [15-17]. However, it is well known that deformation is strongly affected by the exact stress state encountered and that damage and the onset of failure can occur anywhere in the material [18]. It is, in particular, also not limited to the surface. To understand the underlying processes in detail, we need to look at the evolution of damage across the scales and capture information in the volume at the same time. As pointed out by Tasan et al.[19], the lack of information about the three-dimensional behaviour of the damage sites is one of the main sources of uncertainties in present studies.

Here, we therefore set out to go beyond covering a large two-dimensional area [11, 12] or small three-dimensional volumes [20-22], and extend our experimental arsenal to include acquisition and - more importantly - analysis of three-dimensional information for large 3D datasets. In the following, we first summarise current knowledge on damage mechanisms in dual phase steels and available methods for three-dimensional characterisation before outlining our experimental and analytical workflow. Finally, we discuss the results and insights obtained from our three-dimensional analyses as well as the implications for the underlying mechanisms causing the damage.

## Damage Mechanisms in Dual Phase Steels

Different mechanisms have been addressed in the literature for the initiation of deformation induced damage sites in structural materials, which are mostly decohesion at the internal interfaces or fracture of the brittle phase [4, 9-11]. Damage in dual phase steels therefore results from an interplay of different deformation mechanisms at the microscale [19]. The susceptibility of DP steels to damage and failure is highly dependent on the local microstructural features, such as the martensite and ferrite phase fractions [23-25], the carbon content [26], internal structure [27] and hardness of martensite [28], strain hardening capability of the ferrite [29], and the morphology of the phases [19, 30, 31] - besides the amount and directions of strain applied during experimental conditions. Although identification of the exact role of each mechanism is quite a challenging task, these mechanisms have been extensively investigated in the literature based on two-dimensional analysis [3, 4, 11, 12, 24]. Avramovic et al. [32] have reported the decohesion of ferrite-martensite interfaces as the most detrimental factor whereas Stevenson, Mazinani et al. [33, 34] concluded that damage sites initiate inside the brittle martensite phase. Kusche et al. [11] concluded, from their first work using large scale panoramic imaging at different strain levels and for two different microstructures, that the dominant mechanism is likely a function of both microstructure and strain, with a commercial DP800 steel showing a transition from martensite cracking dominated damage formation at small strains to greater importance of interface related mechanisms towards larger strains.

In terms of the local mechanical properties, Ghassemi-Armakani et al. [35] observed heterogeneous hardness values inside individual ferrite grains with higher values in the vicinity of martensite-ferrite interfaces, presumably related to the pile-up of dislocations at the ferrite phase close to the martensite due to the of (geometrically necessary) dislocations [36]. Combining experimental data with finite element method (FEM) analysis, Hoefnagels et al. also discussed the triggering order of active damage mechanisms under different strain paths [9]. By detailed electron channelling contrast imaging (ECCI) measurements coupled with 4 different damage initiation models, recently Liu. et al. [37] proposed that damage initiation is governed by martensite substructure boundary sliding and that the influence of other discussed effects, such as of martensite - ferrite interfacial morphology, phase contrast and residual stress/strain effects, are relatively small.

Since, essentially, "dual phase steel" is an umbrella term for a wide range of materials with varying density and morphology of martensite islands, as well as mechanical contrast between ferrite and martensite, a wide range and even seemingly contradicting results can be expected by different research groups, as each group starts from a sample with different intrinsic composition, microstructure, and properties. The aim of this paper is, therefore, to develop methods that allow us to characterise the response of the material to mechanical stress in terms of the fundamental properties of the sample

Building on the damage mechanisms most commonly observed in the literature, we focus our analysis on the following mechanisms of damage formation and evolution: (i) deformation induced damage by martensite fracture, (ii) deformation induced damage initiated by decohesion of a ferrite-martensite interface, (iii) inclusions originating from sheet processing to avoid their convolution with damage dependent on strain only and (iv) damage at sites exceeding our resolution limit in terms of assigning a clear mechanism of nucleation or growth. We refer to the latter as "notch effect" as these damage sites commonly occur where very fine martensite structures are found that would tend to lead to a stress concentration and therefore large local deformation.

## 3D Damage Analysis

General microstructure-based analyses in materials science are mostly based on the information collected from two dimensions, typically in the form of electron micrographs of a few selected areas of interest [38]. To obtain a more detailed understanding, we need to extend this conventional approach in two ways: First, we have to investigate a large area of the experimental sample at high resolution to be able to generalise the conclusions which one might draw from looking at specific areas of interest. This is particularly important here as the nucleation and evolution of damage is essentially a stochastic process in terms of observation, as damage sites are not known a priori, and a large number of damage sites need to be analysed to be able to draw statistically significant conclusions from the data. Furthermore, the manufacturing process has a strong impact on the microstructure of dual phase sheet samples. For example, rolling the steel leads to large bands of martensite in the microstructure [16], which in turn affects how different regions of the sample respond to external strain. Choosing a few select areas of around 100 $\mu m^2$, as commonly used in the literature, is unlikely to be representative of the experimental specimen as a whole.

Next, we need to include a three-dimensional analysis as most of the physical phenomena, such as the formation of damage, are not constrained to a two-dimensional plane. Usually, the required characterisation of three-dimensional volumes remains out of reach of most conventional laboratory scale equipment. This is particularly true where information is needed to bridge the length scales from the microstructure at the (sub)micrometre level to the gradients within a sample at the millimetre level, as resolution and observable volume tend to scale. Several approaches exist to this end: In atom probe tomography, we can map the three-dimensional location and elemental species of each atom or molecular group, but this method is limited to very small experimental volumes and does not allow for a large-scale analysis [39]. Similarly, tomography by slice and view using focused ion beam (FIB) milling is typically limited to small areas as the related cost typically prohibits the analysis of large experimental samples [40] [41] , even where the use of Xe plasma instead of Ga FIB is becoming possible. This is even more pronounced for X-ray tomography[42] and synchrotron laminography [43]: Apart from the experimental overhead to acquire beam-time at large accelerator facilities, these methods are limited in resolution and therefore not ideally suitable for detailed understanding of the three-dimensional behaviour of damage evolution to the sub-micro scale. Furthermore, some of these methods, such as synchrotron laminography, are not able to distinguish compositionally and crystallographically similar phases of the microstructure, such as, in case of dual-phase steel, martensite and ferrite [44].

Therefore, we explore an alternative approach in this work to analyse the three-dimensional behaviour at large scale and high resolution:  We repeatedly acquire a panoramic electron micrograph and then remove a thin layer of the sample. This has the advantage of being able to analyse large areas at high resolution in x-y-direction, with the resolution in z-direction only limited by the experimental setup for ablating a defined slice of the sample. This approach has the benefit that we are not only able to capture damage sites with high resolution across an area of the order of 1 mm² but can also analyse the damage as the local microstructure changes across the depth of the experimental sample to a depth of several martensite island thicknesses. Furthermore, a combination of polishing and etching allows us to separate the phases of the local microstructure by fast electron microscopy, information which is not accessible by many other methods as indicated above or requires more time intensive electron microscopy methods, such as electron backscatter diffraction (EBSD).

Ultimately, we aim to statistically investigate and understand the three-dimensional hierarchy of damage sites formed inside dual phase steel samples at different strains. To this end, we aim to establish whether the damage observed in the sample after applying stress can be explained by a dominant and more fundamental damage mechanism accessible to 3D analysis compared to the interpretation of 2D images alone. As part of this endeavour, we will investigate the damage events' characteristics in terms of their 3D environment and the likely mechanisms of initiation and evolution. We also consider the differences in interpretation of the dominant mechanism in a comparison of data analysis using the full 3D or 2D data only in order to separate our physical insights from possible artefacts encountered from slicing the microstructure in post-mortem 2D analyses.

# Experimental Setup and Methodology

## Experimental Sample and Preparation

In this study, a commercial DP800 dual-phase steel (ThyssenKrupp Steel Europe AG) was used, i.e., a dual phase steel with a tensile strength of the order of 800 MPa. Three samples were cut out of a 1.5 mm thick sheet metal into dog bone shaped tensile specimens (Figure 1) and then elongated uniaxially in rolling direction (specified by the manufacturer). The tensile tests performed using a microtensile stage (Proxima 100; MicroMecha SAS, France) with a gauge length of the tensile samples of 3.65 mm, a square cross-section of 1.5 mm and applied elongations of 14%, 19% and 25%. After testing, the samples were cut to the gauge length with an electron discharge machine.

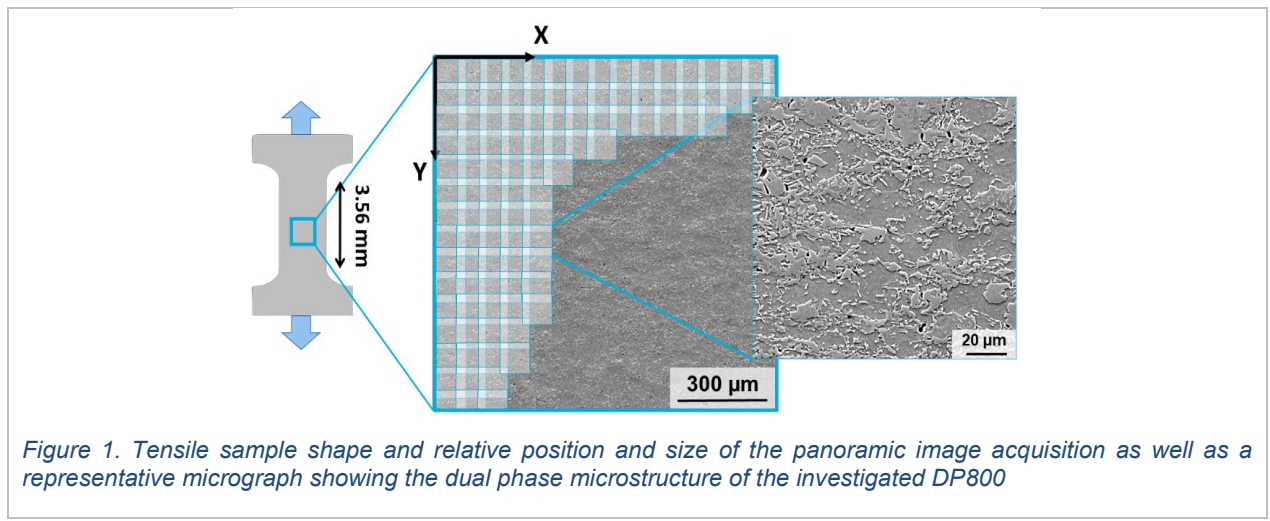

*Figure 1. Tensile sample shape and relative position and size of the panoramic image acquisition as well as a representative micrograph showing the dual phase microstructure of the investigated DP800*

As we wanted to trace the behaviour of individual damage sites across the sample thickness by successive slicing and viewing, we first needed to establish a relationship between the mechanical procedure and the thickness of the ablated material removed before we acquire an electron micrograph of a deeper layer. Using a separate sample of the same DP800 steel, we indented the surface of the sample with different indentation depths starting from 50 nm to 600 nm using a nanoindentation machine (iNano, Nanomechanics Inc.,USA). Then, we devised several combinations of fine polishing and etching steps, and verified the material removed by the various procedures using Atomic Force Microscopy (XE-70 Park Scientific Instruments (PSIA), USA), as well as 3D laser scanning microscopy (VK-X3000, Keyence, Japan). We found that:

- 60 s OPS polishing followed by a cleaning step resulted in 400 nm of material ablation

- 15 s OPS polishing followed by a cleaning step resulted in 100 nm of material ablation, and
- 10 s of light etching with a 1% Nital solution resulted in 150 nm of material ablation.

Here, OPS refers to colloidal silica suspension and the cleaning step was done using a fine grid cloth using the OPS solution.

In order to prepare each of the three samples for 3D analysis, we prepared the initial surface by grinding the surface manually using sandpapers starting from 800 to 4000 grit. Afterwards, the samples were polished mechanically with 6 µm, 3 µm, and 1 µm diamond suspension with an alcohol-based lubricant (99.5% Ethanol + 0.5% Polyethylene Glycol), and we subsequently used OPS for the final polishing and cleaning. Finally, each sample was subjected to a 1% Nital solution for 10 s, which leads to a visible phase contrast between ferrite and martensite in the electron microscope due to preferential etching of the ferrite phase. We refer to this step as 'basic metallographic preparation' in the context of this work. In Figure 1, we give an example of a panoramic image collected from the metallographically prepared microstructure of the sample after uniaxial elongation. Within the next ablating steps, either 60 s or 10 s polishing followed by cleaning and etching was performed, which we refer to as 'fine polishing' in the following. Both processes are visualised as flow charts in Figure 2 with the further details of the 3D data acquisition process described below. To align the electron micrographs across the material slices in the later analysis, we further used a macro indentation machine to create fiducial markers.

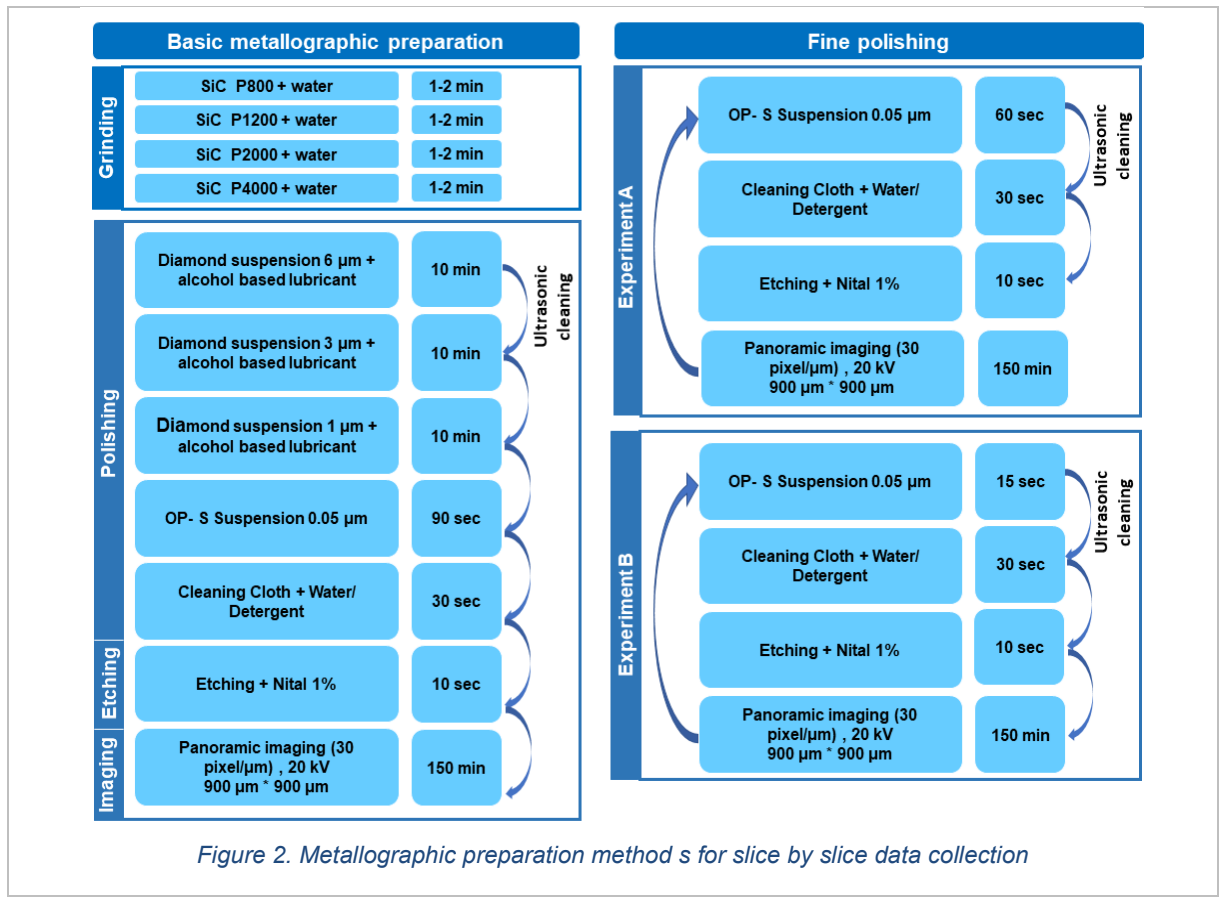

*Figure 2. Metallographic preparation method s for slice by slice data collection*

To capture the three-dimensional structure of the damage sites, we then ablated each sample according to the combination of OPS polishing and etching steps described above, before acquiring a new electron micrograph for the respective slice. We followed this procedure until we obtained a stack of 10 successive slices for each sample.

We chose the depth of each slice in the following way: In experiment A, we used the specimen subjected to 19% elongation, i.e., the intermediate strain level, and ablated the material between the successive images such that 500-600 nm of material was removed. In experiment B (samples with 14% and 25% elongation), we reduced the thickness of each slice to 200-300 nm to be able to track fine structures as we go from the surface deeper into the material (z-direction). This is particularly important for damage sites with a small volume or extension in z-direction (most commonly those referred to as 'notch effect'), as we saw in a preliminary analysis of experiment A that these damage types are at or below the resolution limit of the experiment. Reducing the thickness of the slices further would introduce too many artefacts and was, therefore, not done here: First, more surface artefacts are introduced due to insufficient flattening of the surface when the polishing time becomes too short and, second, the etching procedure required to obtain sufficient contrast for the microstructure to be visible in the electron micrograph would either approach the slice thickness or become too short as well, leading to unsuitable images for the subsequent analysis. The full preparation routine is summarised in Figure 2.

Essentially, after image acquisition (described below), we have two sets of images: In experiment A (for the sample with 19% elongation), we cover a larger volume in z-direction as the 10 slices are 500-600 nm apart, whereas in experiment B (for the samples with 14% and 25% elongation), we obtain a finer resolution in z-direction as the 10 slices are only 200-300 nm apart as indicated in Figure 3.

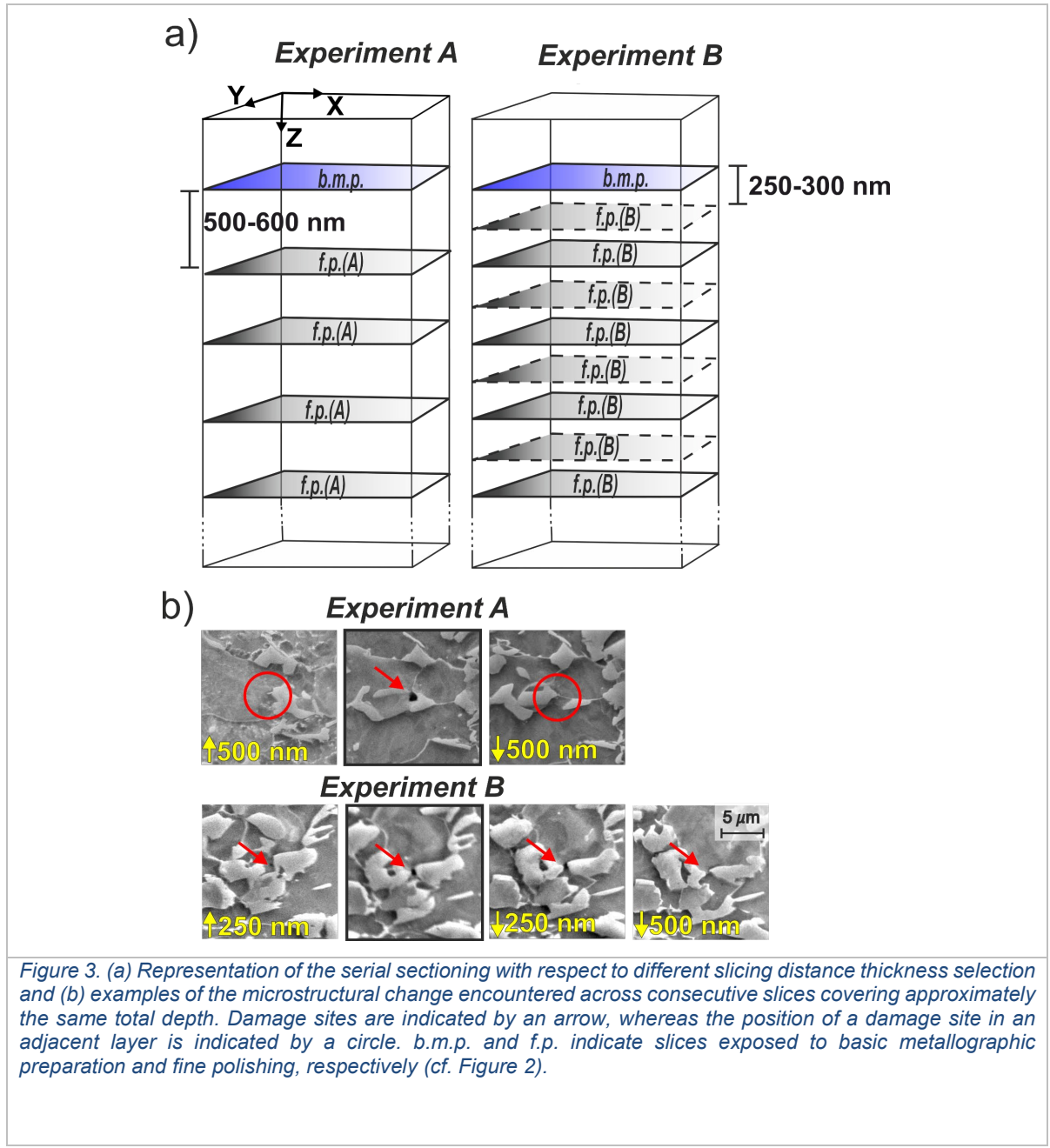

Figure 3. (a) Representation of the serial sectioning with respect to different slicing distance thickness selection and (b) examples of the microstructural change encountered across consecutive slices covering approximately the same total depth. Damage sites are indicated by an arrow, whereas the position of a damage site in an adjacent layer is indicated by a circle. b.m.p. and f.p. indicate slices exposed to basic metallographic preparation and fine polishing, respectively (cf. Figure 2).

## Image Data Acquisition and Pre-processing

We acquired electron micrographs using secondary electron detection in field emission scanning electron microscopes (LEO 1530; Carl Zeiss Microscopy GmbH, Jena, Germany and CLARA, Tecan GmbH, Germany). The spatial resolution of the obtained images was 32.5 nm per pixel in both microscopes. To collect high resolution images from large areas, we recorded a sequence of individual images as squares with a length of 100 µm, and used the Image Composite Editor 2.0.3.0 to obtain a panoramic image with the final dimensions of 850 µm * 850 µm.

A typical example of an electron micrograph showing the relevant damage classes used in the image analysis is shown in Figure 4.

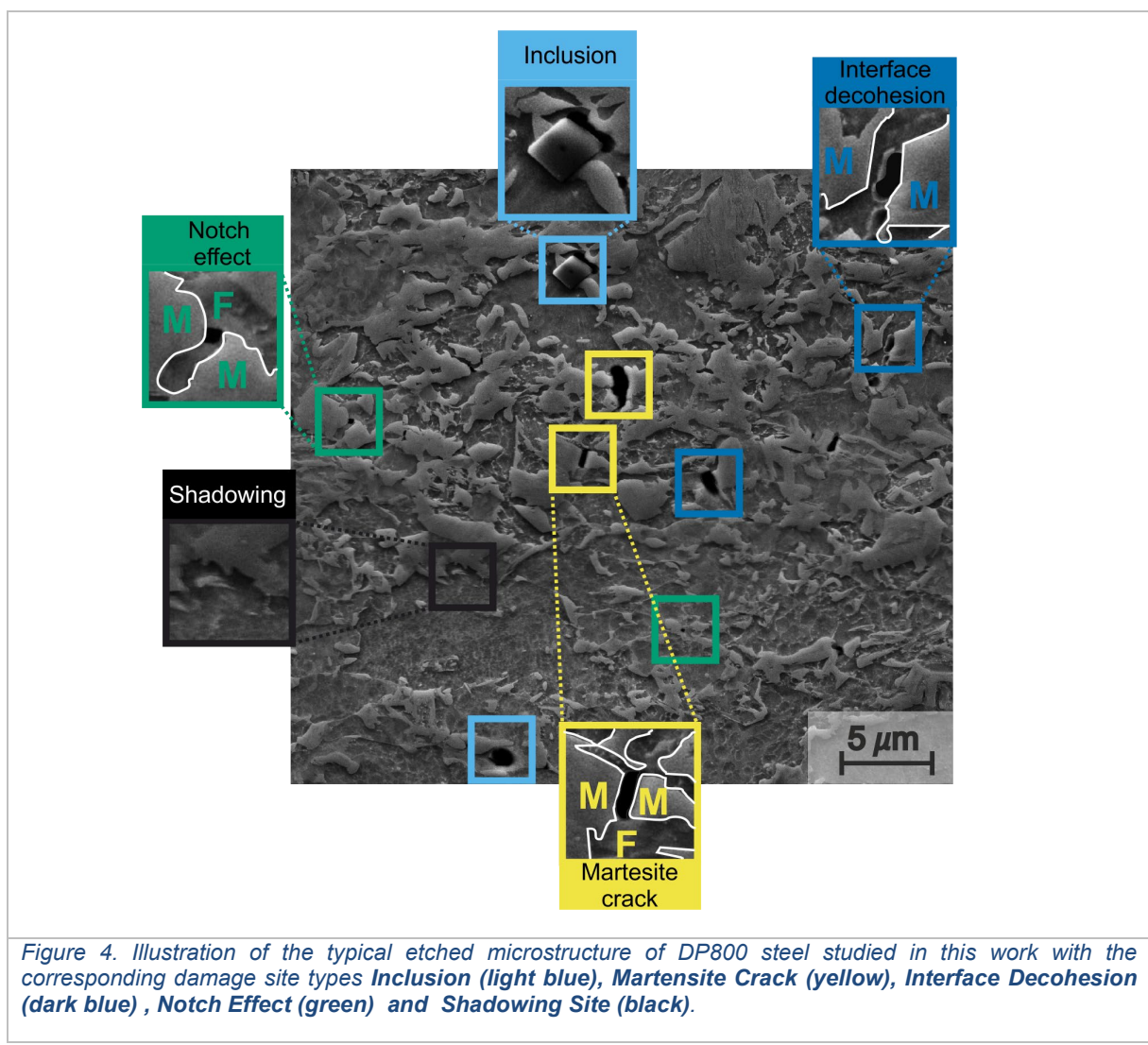

*Figure 4. Illustration of the typical etched microstructure of DP800 steel studied in this work with the corresponding damage site types **Inclusion (light blue), Martensite Crack (yellow), Interface Decohesion (dark blue), Notch Effect (green) and Shadowing Site (black)**.*

Since the panoramic image is created from many individual images, the resulting panorama shows variations of brightness and contrast across the image. These are due to variations in brightness during image acquisition. In order to compensate for this effect, we apply a pseudo flat field correction [45, 46]. An example of the improvement obtained with this approach is shown in the Supplementary Materials.

The stack of 10 successive panoramic images is roughly of the same experimental area as defined by the fiducial markers. However, since the sample has to be removed from the microscope between each image to ablate, polish and etch the material, the sample will not be in exactly the same position for each image acquisition. Therefore, we need to align the successive images such that we can track the evolution of damage sites throughout the volume. This allows us to identify which damage site in the current slice is associated with which damage site in either the successive or previous slide, or if a new damage site emerges in the current slice. The alignment of images is also called image registration. To achieve the necessary alignment, we adopted the registration strategy for image series proposed in [47], i.e., each image is aligned to the next and the resulting alignments are combined to align all images in the stack to the first image in the stack. As transformation model, we used an affine coordinate transform that can correct translation, rotation, tilt, and shear. Each image pair is aligned using a coarse-to-fine strategy, i.e., the alignment is first done on a coarse image resolution and the result is used as initialization for the next finer resolution. For a given resolution, the alignment is done by minimizing the sum of squared difference between the images to be aligned over the affine transformation of the image that is deformed. This is modelled as non-linear least squares problem and solved using the Trust Region Reflective algorithm. Due to the very high resolution of the images, the resolution refinement is stopped before the full resolution is reached.

In the subsequent three-dimensional analysis of the damage sites, we focus on damage sites that are fully contained within the analysed volume and, in particular, do not take damage sites into account that are already present in the initial surface layer. This is done for the following two reasons: First, since we have no experimental data beyond the initial surface layer in the upwards (negative) z-direction, we would not be able to determine the potential type of damage along the entire depth of a given site. Second, as seen later in Figure 8, the number of observed damage sites in the initial surface layer is indeed significantly higher compared to all subsequent layers. This is the only layer subjected to the initial basic metallurgical preparation, and to avoid any potential bias we do not use the damage information from the surface in the subsequent analysis. Figure 5 illustrates a range of typical topologies we consider in our analysis. Topologies that we do not consider in this work are shown as hashed objects.

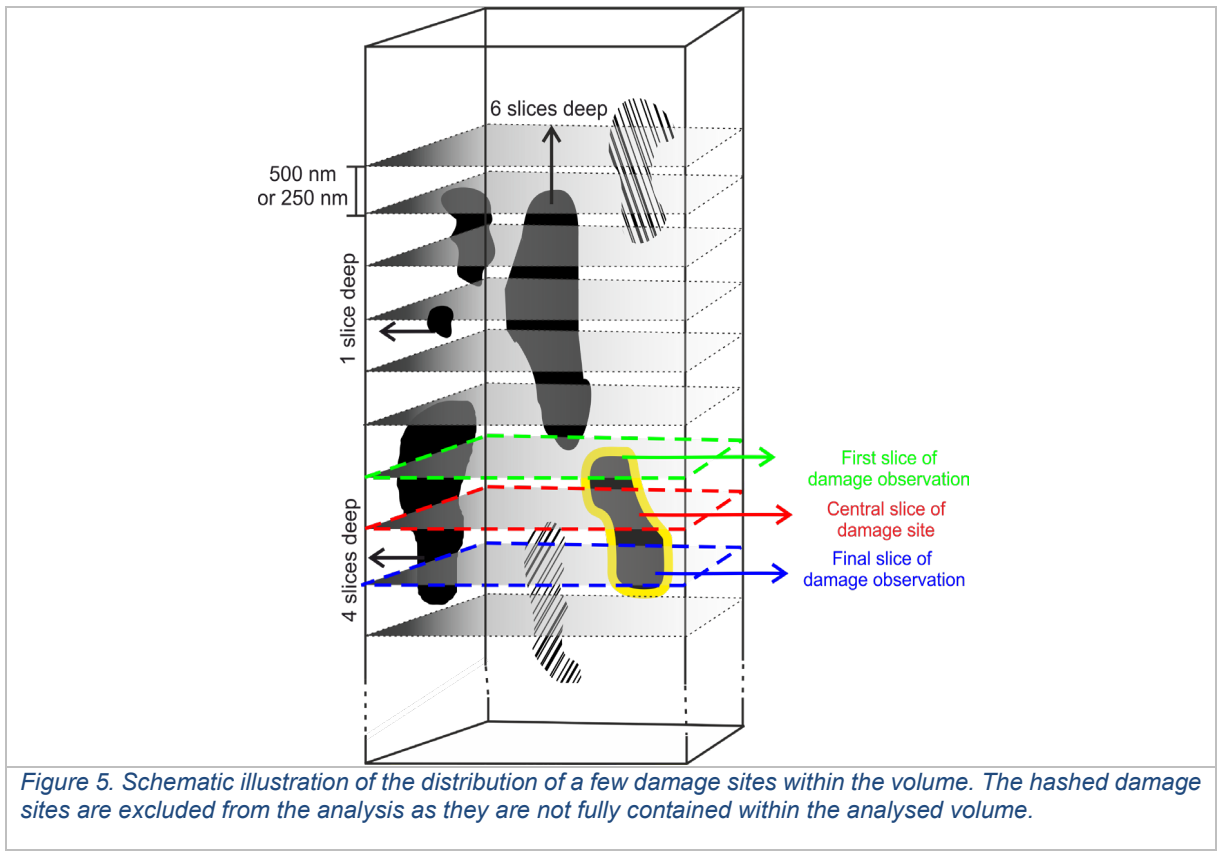

*Figure 5. Schematic illustration of the distribution of a few damage sites within the volume. The hashed damage sites are excluded from the analysis as they are not fully contained within the analysed volume.*

## Damage Site Analysis in 3D

We follow the procedure established in [11] for the detection and identification of damage sites in each recorded slice of the material for each of the three samples with 14%, 19%, and 25% elongation, respectively. Essentially, each potential damage site appears as a black region in the panoramic electron micrograph. Damage site candidates are first identified using the DBSCAN clustering algorithm [48]. Then, a threshold is applied to remove small artefacts and only images with greyscale values exceeding this threshold are kept and extracted from the panoramic image. These individual images are then classified using a sequence of two convolutional neural networks. First, we identify inclusions, where a foreign particle used to be stuck in the local microstructure and may then have fallen out of the sample during the preparation process or not. These inclusions are notably different from the other damage classes regarding size and shape and are therefore identified first. Furthermore, inclusions are not deformation induced damage and are therefore of limited or, where they may be found to induce damage formation, separate interest in understanding the response to the material to an applied stress. In a second classification step, we then use a second neural network to distinguish between the physical damage classes: martensite cracking, interface decohesion and notch effect. Further, we also include image artefacts such as shadows in the list of categories to improve the accuracy of the algorithm [12]. An example of each class is shown in Figure 4.

In the next step, we relate the now identified and classified damage sites across the stack of images obtained from the slices of the experimental sample. In the simplest case, where all slices are aligned perfectly by the registration procedure and each damage could be described by a circle in a homogeneous material, we would find that the centres of each circle have the same (x,y) coordinates and only differ in z-direction across the slices or stack of images. However, in a real sample, we have to account for the following effects: First, the material is not necessarily homogeneous but has a distinct microstructure. This in turn can influence where damage occurs and how it evolves through the depth of the material. Then, depending on the type of damage and the local microstructure, the shape of the damage is not necessarily circular. As the shape changes with the evolution of the damage site and throughout the stack of images, the centre of gravity of each damage site will also necessarily shift across the slices. Additionally, new damage sites can emerge in any of the slices. Finally, since the registration procedure is a numerical optimisation, small uncertainties in the procedures are to be expected. To account for these effects, we apply the following procedure: As the occurrence of damage is relatively rare, even in the sample elongated by 25%, the individual damage sites are relatively far apart from each other. This means that there are no multiple damage sites in close proximity to each other, and we can therefore treat any damage site found in adjacent slides as related to the damage in the current slice if they are found within a certain search radius. This radius is defined by the registration procedure and given by the effective shift in (x,y) coordinates between two adjacent slices: $S = \sqrt{(\Delta x)^2 + (\Delta y)^2}$ where $\Delta x$ and $\Delta y$ are the effective shift in coordinates of the two adjacent slices, namely the Euclidean distance. For example, if the effective shift is -10 pixels in x-direction and +5 pixels in y-direction, the search radius for a potential damage site in the adjacent slice would be r=11 pixels as illustrated in Figure 6.

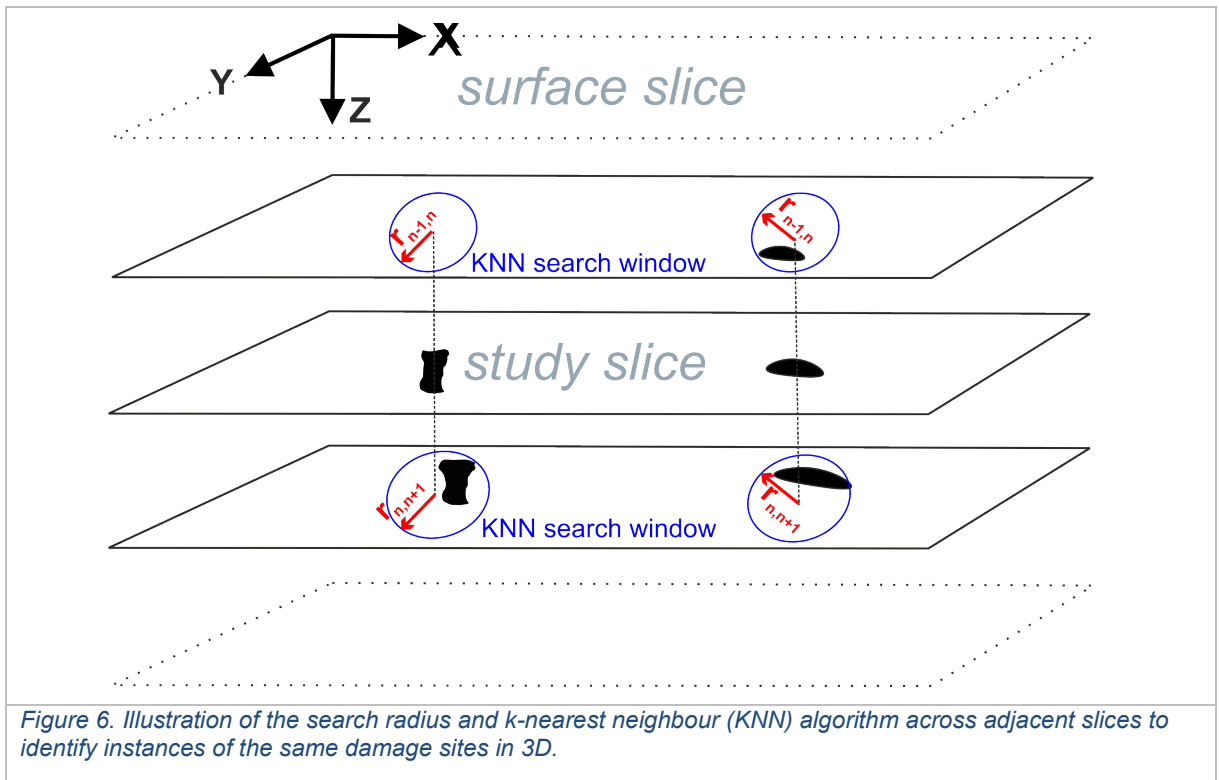

*Figure 6. Illustration of the search radius and k-nearest neighbour (KNN) algorithm across adjacent slices to identify instances of the same damage sites in 3D.*

To determine whether or not a given damage site extends to the adjacent slices, we use a k-nearest neighbour algorithm [49], where the free parameter is set to k=1 as we look for exactly one potential damage site in adjacent slices within the search radius r. If a damage site can be found within this search radius, we attribute this damage site to the damage volume in question. If we do not find any damage site, we conclude that the damage either begins or ends in the current slice. For example, we start with the slice n shown in Figure 6 and look for a damage candidate in slice n-1 and n+1. In this case, we do not find a damage site in the search window in slice n-1 in the left case, but we do in slice n+1 for both damage sites shown. Therefore, we conclude that the left damage first occurs in slice n and then extends to slice n+1 and possibly beyond as we repeat the procedure for the further slices in the stack.

A concrete example of this procedure is shown in Figure 7, again for two cases. In the top row (a) we identify a damage site in the first slice with coordinates $x_0$ and $y_0$ and then follow the procedure across the next three slices (↓x1,↓x2, ↓x3). Here we can identify the damage across the slices within the search radius r. Note that an additional damage occurs only in Figure 7.a) in slice ↓x3 at the bottom right. However, this damage is outside the search radius r for the current damage site under consideration and is therefore not attributed to the current damage site. The bottom row (b) illustrates the case where we identify a damage site in the second slice (x1) that extends to the two subsequent slices ↓x2 and ↓x3, but is not visible on the surface, meaning that this is a new damage site that emerges in slice x1.

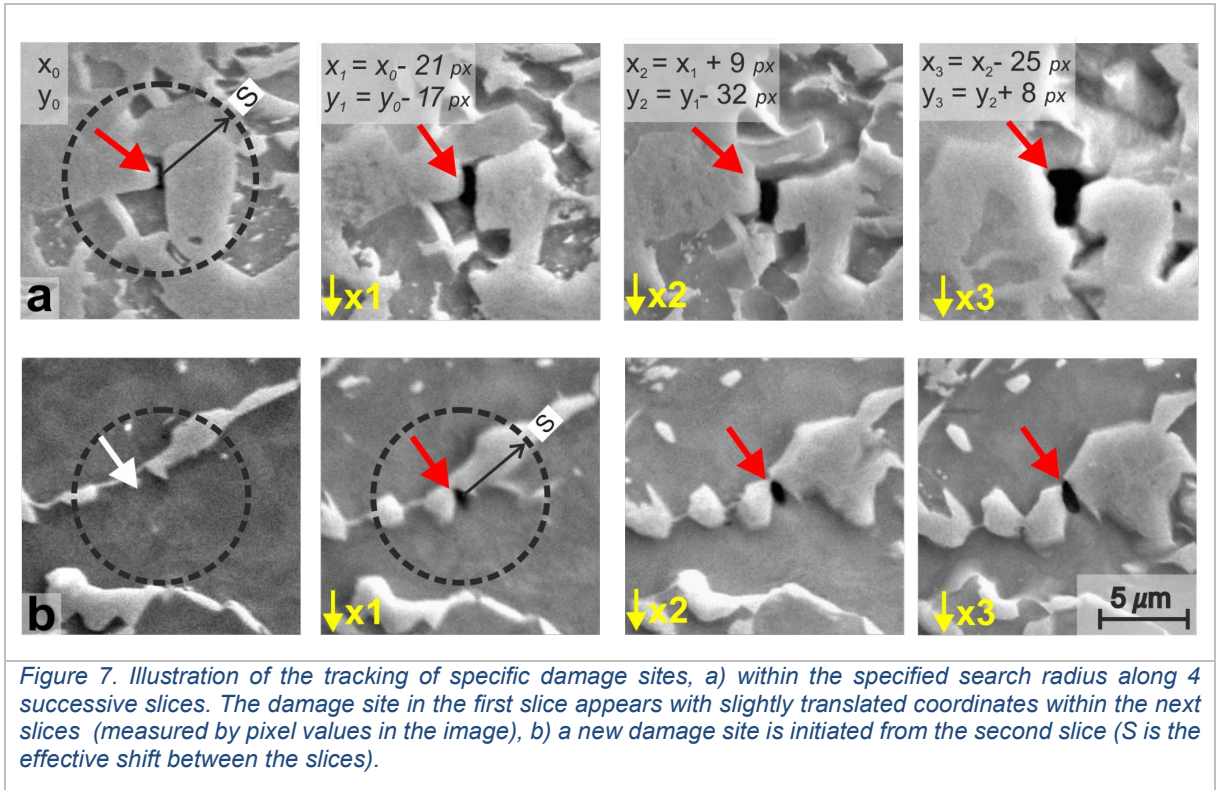

Figure 7. Illustration of the tracking of specific damage sites, a) within the specified search radius along 4 successive slices. The damage site in the first slice appears with slightly translated coordinates within the next slices (measured by pixel values in the image), b) a new damage site is initiated from the second slice (S is the effective shift between the slices).

Completing this procedure for all three samples, we obtain a set of three-dimensional damage sites across the volume for each sample. For each damage volume, we then know the (x,y) position of the damage sites in each slice, the damage type of each damage site attributed in each slice, and how each damage site is either connected to a damage site in adjacent slices, or if the damage site occurs in a given slice for the first time.

# Results

## Distribution of damage types within 2D slices

First, we investigate the number of sites of each type of damage per slice separately for each of the three samples. This is visualised in Figure 8, where each damage type is indicated by the following colours: martensite cracking (MC) in yellow, interface decohesion (IF) in dark blue, notch effect (NE) in green, inclusion (INC) in light blue and "not classified" (NC) damage sites in grey. The case "not classified" refers to the cases where a damage site candidate has been identified but the algorithm did not attribute this candidate to either of the physical damage classes (martensite cracking, notch effect, interface decohesion, inclusion) or imaging artefact (shadowing).

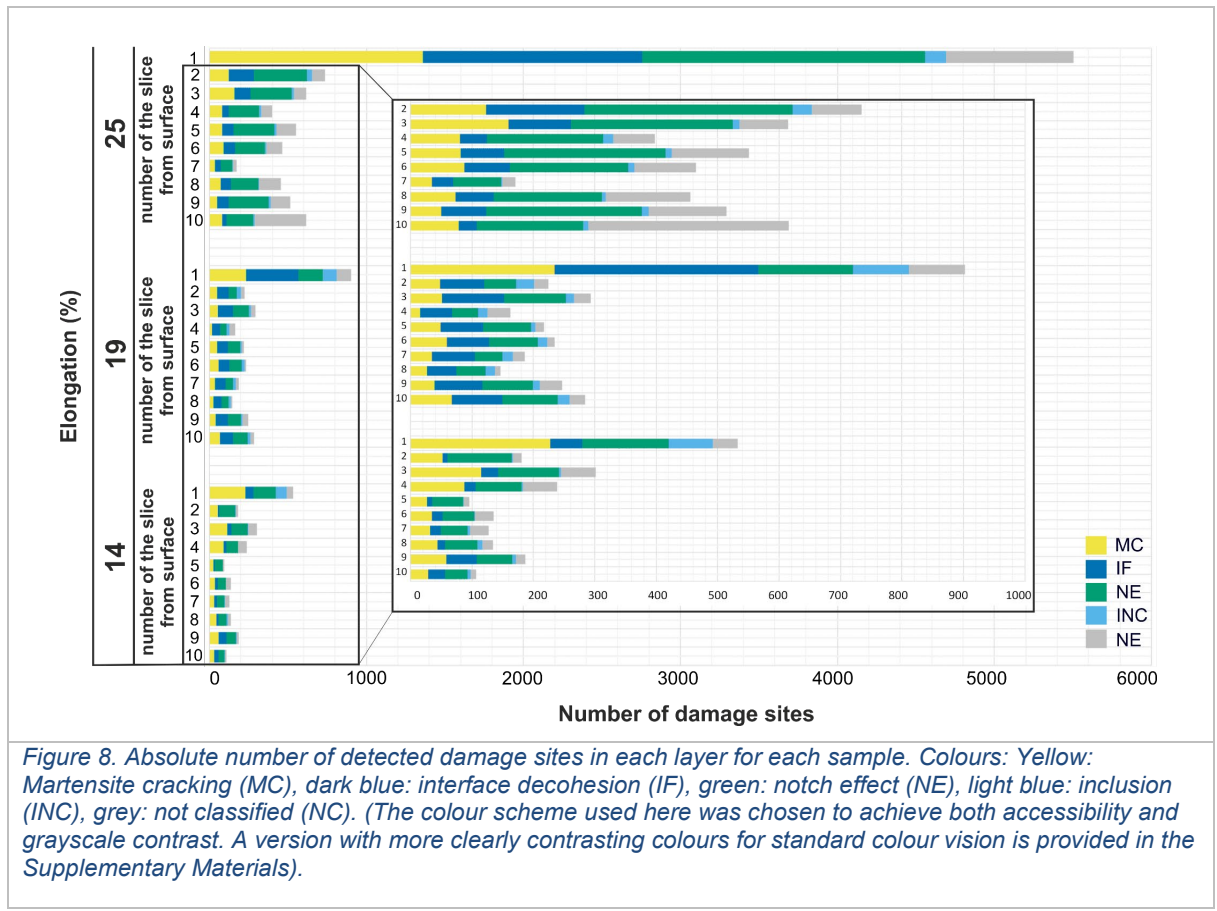

*Figure 8. Absolute number of detected damage sites in each layer for each sample. Colours: Yellow: Martensite cracking (MC), dark blue: interface decohesion (IF), green: notch effect (NE), light blue: inclusion (INC), grey: not classified (NC). (The colour scheme used here was chosen to achieve both accessibility and grayscale contrast. A version with more clearly contrasting colours for standard colour vision is provided in the Supplementary Materials).*

At first glance, we notice that the number of damage sites in the surface layer is considerably higher compared to all subsequent slices, independent of applied strain. Most of this effect is likely due to the initial, more aggressive basic metallographic preparation step of the sample and therefore an artefact of the handling of the sample, as we will discuss later. To avoid unintended bias, the damage sites found in this slice are therefore not used in the later analysis.

In terms of the evolution of the number of damage sites with elongation, we notice that the number of damage sites follows our intuitive expectation and previous experience [11], namely that the sample exposed to the least amount of stress and strain (14% elongation) shows the least damage, whereas the sample elongated by 25% has the highest occurrence of damage. The number of inclusions is higher in the sample subjected to 19% strain compared to the other two samples. We believe this is due to the inherent properties of that sample and must be unrelated to the applied strain. We therefore included this data in our analysis of the three-dimensional distribution of damage at each site associated with inclusions but will exclude the quantitative contribution of inclusions to the evolution of damage sites in most of our analysis in order to separate the deformation induced damage sites in the focus of this work.

## Depth and depth distribution of damage sites

Next, we look at the occurrence of damage for the cases that a damage site extends over two, three or more slices. Overall, we find that the more slices the damage sites extend over, the smaller their number, i.e., most damage sites are two slices deep, followed by damage sites three slices deep, etc., as illustrated in Figure 9. The data for all three experiments are very well represented by an exponential fit, that is using a function of the form $f(x) = a \cdot exp(b\ \text{x})$, where $a$ and $b$ are the fit parameters and $x$ the damage site depth, as shown in Figure 9.

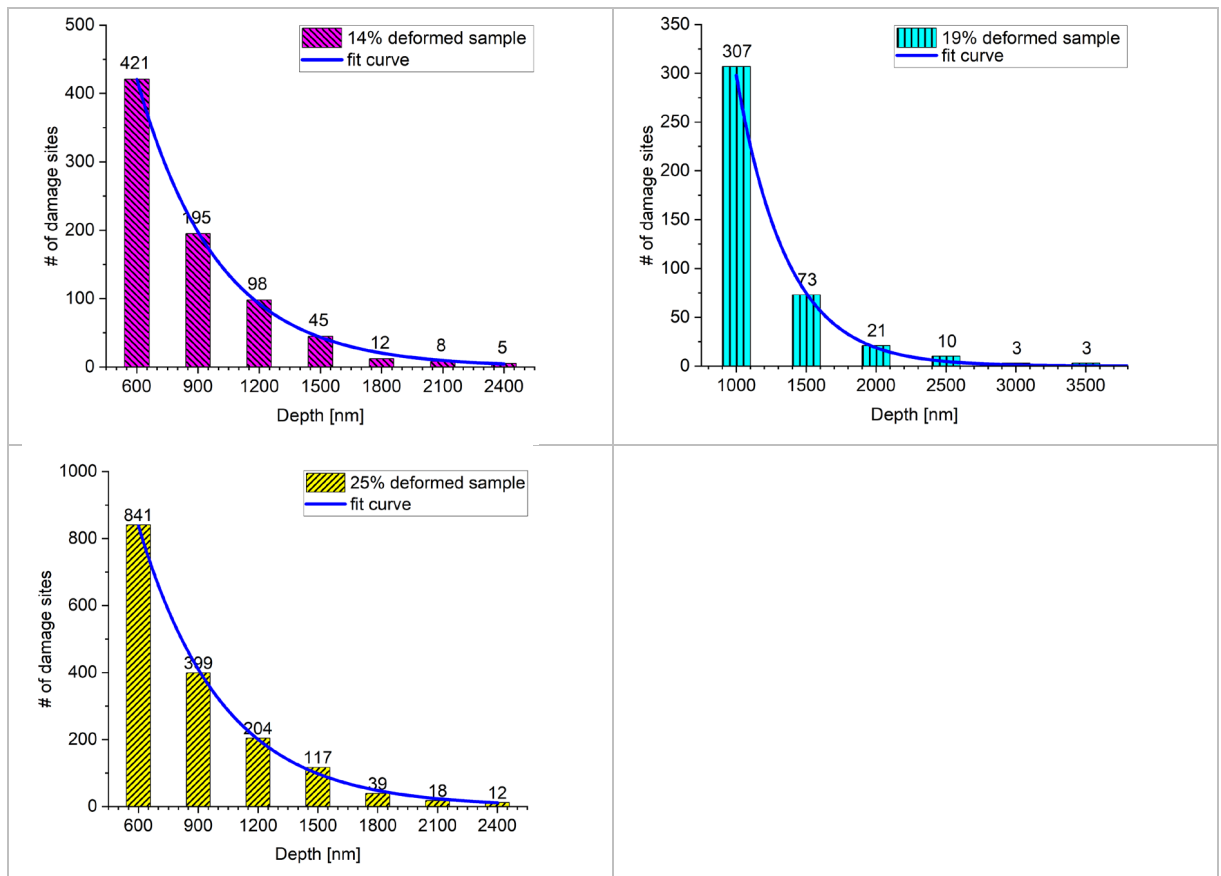

*Figure 9. Histogram of the number of damage sites along the depth in the samples elongated by a) 14%- Experiment B, b) 19% - Experiment A, c) 25% - Experiment B. The data can be represented by an exponential fit, as indicated by the blue lines.*

Since we only consider damage sites that are fully contained within the volume we analyse and have excluded the first layer closest to the surface, the maximum depth a damage site can penetrate into the material is eight slices deep. The physical damage sites can, of course, be deeper, but would then no longer be fully contained within our three-dimensional sampling, and therefore not considered in our analysis.

In the above data, we observe a higher number of shallower damage sites overall in the volume as well as a strong decline in the overall number of damage sites with increasing size in slicing direction. We observe the same trend in Figure 10. In this figure, the number of damage sites that exist in any given slice are shown together with the depth to which these damage sites extend. The lighter colour of the short bars highlights the majority of the damage sites occupying up to two slices in depth. In these visualisations the colour scheme always remains the same, but the maximum value differs to account for the different maximum number of sites found for a given depth. In this way, we identify that the distribution of damage sites appears largely homogeneous, i.e., fluctuations are not systematic, with the exception of the larger number of small damage sites towards the surface.

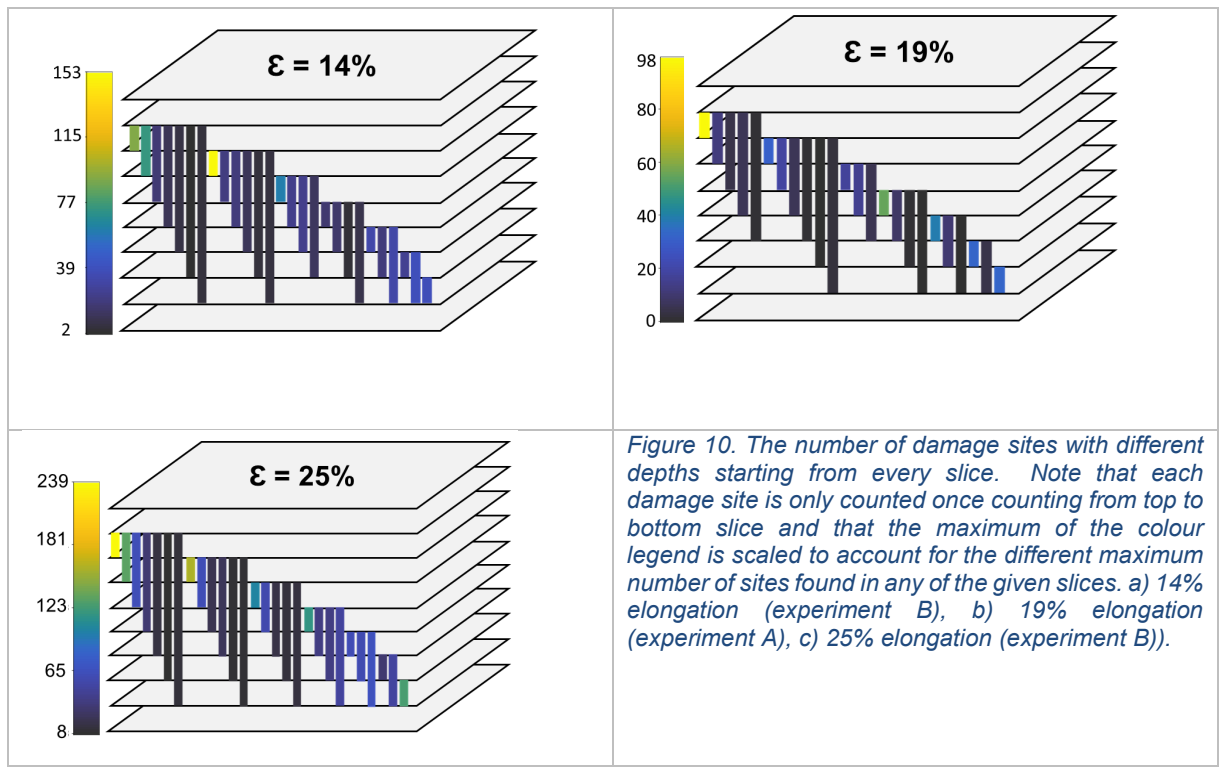

Figure 10. The number of damage sites with different depths starting from every slice. Note that each damage site is only counted once counting from top to bottom slice and that the maximum of the colour legend is scaled to account for the different maximum number of sites found in any of the given slices. a) 14% elongation (experiment B), b) 19% elongation (experiment A), c) 25% elongation (experiment B)).

## 3D characterisation of damage type across slices

A key aspect of this work is to relate the damage sites identified in the individual slices across the stack of images. This allows us to track the type of damage across the depth of the material and analyse how a specific damage site evolves along its depth. In particular, we are interested in whether certain damage types dominate for many sites or whether any changes in apparent damage mechanism can be associated with either a physical origin in the microstructure and underlying deformation mechanism, which in the end, would lead to a better understanding of the data that are usually obtained from single cross-sections.

To visualise the behaviour of the damage across the slices, we use Sankey diagrams. These are well suited for this purpose of visualising the underlying multi-layered and inter-connected data as they not only allow us to illustrate the relative number of the different damage types in each slice, but also how they change between slices. These diagrams, shown in Figure 11, can be interpreted in the following way: The occurrence of a particular damage type in each layer is indicated by a horizontal bar that denotes the slice (excluding the surface slice) in which this damage occurs. The larger the fraction of a given colour on the horizontal line, the more abundant the respective damage type is in this slice. These bars are then connected by vertical lines that indicate the change or continuation of damage type from one slice to the next. The thicker this connection is, the more prevalent the transition between the two damage types is in the adjacent slices. The vertical lines therefore trace the provenance of the damage from the uppermost layer downwards into the deeper slices of the material. The Sankey diagrams do not contain again the information about the absolute number of damage sites as a function of their depth shown in Figure 9 and Fig. 9 and Figure 10. Instead, sites of equal depth are grouped into individual Sankey diagrams, and these are in turn aligned by physical depth in Figure 11, such that the thinner slices give rise to more diagrams. The diagrams for the thicker slices (in the sample elongated by 19%) are given in the supplementary materials. The underlying data is also again resolved by slice in which the damage sites start for the 19% elongated sample.

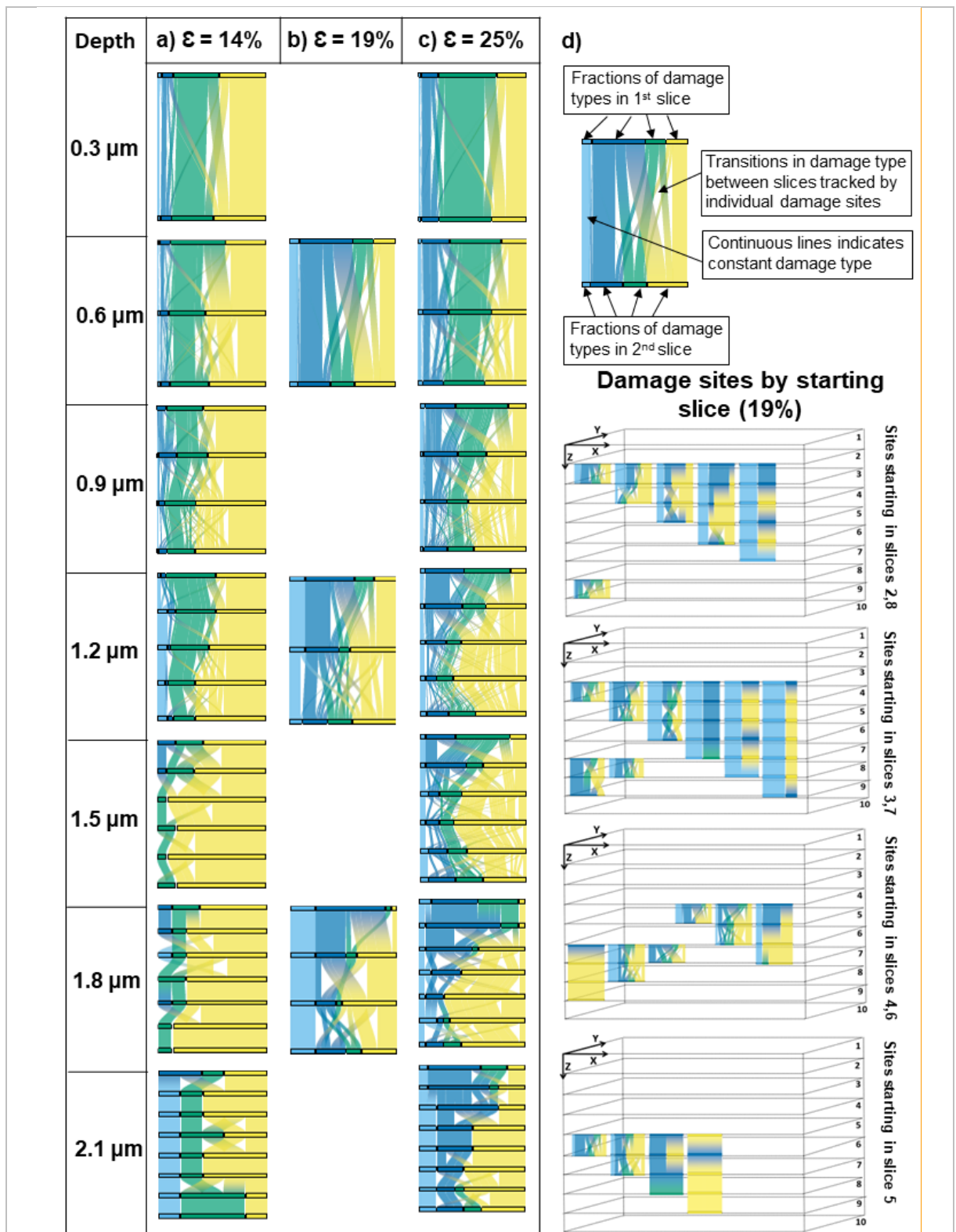

*Figure 11. Sankey diagrams by sample elongation and damage site depth tracking the 2D classification of damage sites in 3D. Colours identify damage type, horizontal bar width indicates type fraction at the given site depth and vertical connections trace damage type evolution along the depth for all damage sites individually. Colours (from left to right): Light blue: inclusion, blue: interface decohesion, green: notch effect, yellow: martensite cracking.*

*(The colour scheme used here was chosen to achieve both accessibility and grayscale contrast. A version with more clearly contrasting colours for standard colour vision is provided in the Supplementary Materials).*

Figure 11.a,b, and c show the corresponding details in Sankey diagrams for the samples subjected to 14%, 19% and 25% strain, respectively. As an example, consider the case of all damage sites in the volume that are two slices deep in the sample subjected to 19% elongation, that is the topmost plot shown in Figure 11.b Each damage type occurs both in the first and in the second slice. The overall fraction of inclusions (light blue) and sites classified as 'notch effect' (green) remain approximately constant. However, while nearly all inclusions preserve their types along the depth, about half the notch sites present as either interface decohesion (dark blue) or martensite cracking (yellow) in the adjacent slices. Interface decohesion is more prevalent at the top of those damage sites which are two slices deep, whereas martensite cracking is more prevalent in the lower half. This is reflected in the connections between the colours indicating the damage type with the thickest connections running from the blue interface decohesion sites to the yellow martensite cracking sites. We will assess the implications of these plots in the discussion but note here that for the small damage sites continuing through only two slices, these plots can then be interpreted as follows: small inclusions do not induce ductile damage vertically in the surrounding slices which are at a distance of approximately 500 nm above and below. This is in contrast to the deformation induced damage, which shows frequent type transitions. Considering the deformation induced damage sites, we see that the microstructure, containing a complex martensite morphology that is far from globular (Figure 1), leads to many instances of 'notch effect' sites. They can be either small in all dimensions or they can be small within the x-y plane in which they were classified only but extend far into the z-dimension. The first is to be expected for the shallowest damage sites associated with thin martensite struts. Accordingly, the majority of these sites remains thin also in the z-direction, reflecting the branched martensite structure.  However, there are also a large number of these sites that are associated with thicker parts of a martensite island above or below. In these cases, the damage mechanism can therefore be identified more clearly as interface decohesion or martensite cracking in the adjacent slice and both of these occur in conjunction with notch sites. As expected, the class 'notch effect' therefore necessarily comprises two sub-classes which cannot be distinguished based on 2D analysis alone: instances of very small damage sites where the martensite becomes very thin and instances that only look small in cross-section, where the slice runs through a thin part of a bigger martensite island.  We also note that there are fewer martensite cracking sites in the top layer than in the bottom layer, which in contrast exhibits more interface decohesion sites. The reason for this may not be deduced from knowledge of the microstructure and the individual diagram alone and must be investigated later after considering the entire dataset.

The following phenomena have been generally observed in Figure 11:

The inclusions, as the material's inherited damage from synthesis and processing, remain mostly isolated, that is they are not usually associated with deformation induced damage above or below, regardless of their size. We also note that the different fractions of inclusion sites, with the sample strained to the intermediate strain of 19% showing the highest value, remain approximately constant in their relative number. That is, inclusions tend to become more frequent towards larger sizes and, in the sample with higher inclusion density, their fraction remains higher throughout.

Comparing the distribution of the damage sites at different strains, that is from left to right in Figure 11a-c, reveals that there is transition from martensite being dominant at small strain to interface decohesion making up an increasing fraction of damage sites with increasing strain.

In looking at the data from top to bottom, which is from shallow to deeper sites extending through several slices, we note that shallower damage sites show a higher fraction of sites classified as notch effects. As the damage sites become deeper, several transitions between different types become more frequent. Considering these transitions, we note two further observations: First, that in general there appears to be an imbalance in the distribution of interface decohesion sites with them becoming more pronounced towards the top of the analysed damage sites and , second, that across those damage sites which extend through several slices, a 'yellow belly' tends to develop, indicating that the centre of the damage sites consists of a cracked martensite island while above and to an apparently smaller extent also below, the martensite cracks extend into interface decohesion sites.

# Discussion

Analysing damage in three dimensions is critical to our understanding of the behaviour of the material under stress. In this work, we set out to provide a method that will enable us to bridge the gap between standard 2D imaging and tomographic analyses. While the first usually misses important information on the underlying microstructure and/or changes the stress state when the surface is cut free, the methods allowing 3D analyses usually suffer from either a lack in resolution or interrogated volume.

Following the imaging, slicing and alignment procedure introduced here with the automated damage classification networks, we find that this method does indeed provide new insights into the prevalent damage mechanisms and will allow more general deductions on the interplay of microstructure morphology and strain path in particular in these highly variable advanced steels.

In the following, we will first discuss the identified distribution of damage site sizes, the proportion of each mechanism identified at different strains and how they relate to the microstructure morphology in the context of the available literature. We will then present examples from the results shown above to illustrate specific cases of damage site evolution into the third dimension, which has now become accessible through the volume, and relate these insights to our additional understanding of the dominant damage mechanisms, their dependence on strain and their interaction in 3D. By directly comparing successive 2D analyses with the information collated as 3D data, we also identified or confirmed several artefacts from sample preparation and imaging, which we will discuss briefly to highlight how they affect the analysis and how they may be avoided or considered more consciously.

## Size distribution of damage sites

We found that the distribution of damage sites according to their size (represented here by their depth) is well represented by an exponential fit (Figure 9), in agreement with previous reports in the literature [13, 50]. The exponential fit would be consistent with a stochastic distribution. Here, this rather implies that the observed microstructure contains no dominant length scale within the range investigated, i.e., between the order of 200 nm and a few microns. A possible length scale would be, for example, a common martensite island diameter that would give rise to many sites of similar depth in case of through-thickness cracks. We note that we do indeed observe a drop in damage sites towards smaller depths of 1 slice (200 nm thick), consistent with this observation and a preliminary analysis of the cross-sectional view but cannot quantify depth further below this threshold.

Previous work on the same material using 2D microscopy identified a similar exponential distribution in terms of the void area of damage sites taken across a range of samples, strains and triaxialities [13]. Due to the deformation to failure and the coalescence of sites associated with the large strains, the distribution also included several damage sites towards very large areas and as only 2D SEM imaging was used, the area extends to very small values of the order of 0.1 µm² below the resolution limit of slicing used here.

In order to assess the significance of potentially excluding sites and validate our findings from the 3D analysis, we re-analysed a cross-sectional view of a sample taken from the same DP800 steel and elongated to failure in tension with the image spanning an equivalent plastic strain of approximately 43 - 53 % from [13]. In this case, we measured the depth of each damage site as the vertical extension (relative to the sheet plane) of each site to correspond to the same physical dimension as analysed from the sliced volume. The cross-sectional analysis of the number of damage sites is shown in Figure 12 An exponential function again fits this data of the number of damage sites as a function of their depth well. More importantly, we also note that towards large damage depths, the number of sites exceeding our analysed depths of the order of 2 - 4 µm, depending on slice thickness, is very small. While the coalesced sites leading to material failure may therefore escape this kind of analysis, for the description of damage evolution during plastic deformation at plastic strains well before failure, as is the intention here, the investigated depth contains a representative distribution of damage sites in terms of their depth. Encountering again an exponential distribution of depths for damage sites is in fact consistent with the particular microstructure present in the DP800 steel, which is characterised by a highly inhomogeneous and branched out martensite distribution without a clearly dominant island size that could give rise to a peak in the damage site depth distribution, e.g. where martensite islands of similar dimensions exhibit through-thickness cracking. In other dual phase microstructures this may well occur and should be easily revealed by a similar analysis.

In either case, whether it relates to measuring the area in the sheet plane using a 2D micrograph, the depth in a 2D cross-sectional image or the depth from a slicing experiment, we note that towards small depths the resolution limit is reached, and the number of detected sites declines. This is the case in the area distribution in [13] as well as the slicing data obtained here, where a depth bin cannot be given for any damage sites extending fewer than two slices and areas are no longer reliably measured where a large fraction of sites may become very thin and escape detection or quantification. Where higher resolution of the damage size distribution is of the essence, e.g., to identify the role of anisotropy in martensite islands due to banding, 2D imaging as well as slicing could be performed from two or three perpendicular directions, such as sheet normal, rolling and transverse directions. However, in most cases it will be sufficient to simply complement a large area slicing analysis from one prioritised direction with either 2D data from a second or third direction (as applied here to confirm our 3D depth distribution measurement) or a higher resolution but smaller volume analysis, such as by FIB slice and view.

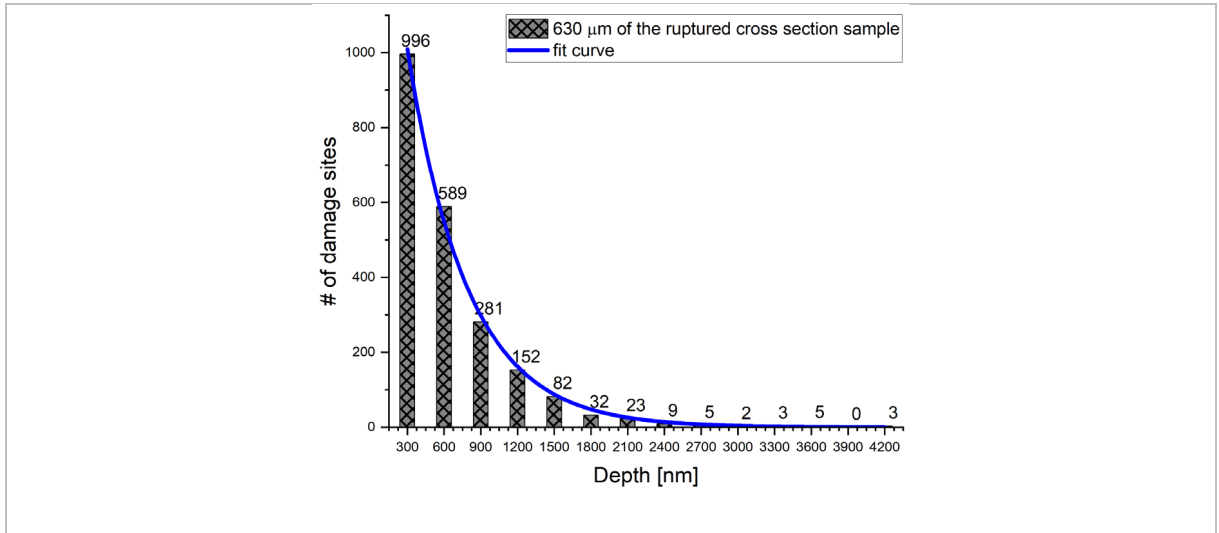

*Figure 12. Histogram of the number of damage sites along the perpendicular cross section of a sample deformed to failure with the depth taken as the vertical damage height, equivalent to the depth in z-slicing-direction.*

## Strain dependence of damage mechanisms analysed in 2D

Comparing the proportion of the identified mechanisms at damage sites in Figure 11, we observe that with increasing strain, the dominant mechanism changes from initial martensite cracking to interface decohesion. This is consistent with previous observations elsewhere [11, 14, 51]. In addition, at the highest strain, a larger fraction of sites is identified as notch effect as a result of localised plastic strain at the ferrite regions between closely spaced martensite islands and thin martensite particles surrounding ferrite matrix, as shown by Ahmadi et al. [30]. In addition, this highlights that, as straining proceeds, further sites of local stress concentration are activated. Overall, this finding is in agreement with the findings of Avramovic-Cingara et al. [32], in that for microstructures with banded martensite, cracking rather than interface decohesion appears to govern damage formation.

There is also a larger number of sites identified as inclusions at the intermediate strain of 19%. In this case, we conclude that while the large area and even the evaluated number of slices appear representative in terms of the captured damage site sizes, their distribution, and active mechanisms of deformation induced damage, the distribution of inclusions is more variable across a sheet of dual phase steel. As inclusions may be introduced as a result of local contamination that is expanded and broken up during rolling, such variation between samples is not surprising. The method presented here is therefore not suitable to assess inclusion distribution on this much larger scale. However, as we find that most inclusions tend to be large, light microscopy, indirect detection of the introduced elements or any secondary roughness and, if need be, 2D panoramic electron microscopy would suffice.

In summary, panoramic 2D microscopy extending to the order of 1 mm² or, considering that many damage sites are small and therefore likely independent, tens of mm² acquired across the 30 panoramic images presented here, reveals consistent information on active damage mechanisms in 2D views and their dependence on strain. In addition, the number density of damage sites evolves consistently across different samples, strains, slices and viewing directions, following an exponential distribution with sharply increasing number densities towards smaller sites up to the relevant resolution limit. Deformation induced damage is therefore well represented in this way. However, the investigated area is found to be insufficient to exclude variations in inclusion density.

## From 2D to 3D view

A major shortcoming of 2D analyses is that the information about the local stress state stemming from the dual phase structure remains unknown. The single polished surface therefore does not yield information about the interaction of mechanisms, the evolution of sites along the depth within the dual phase microstructure and whether the shifts in dominant mechanism observed in the 2D slices at different strains originate from the evolution or exhaustion of the martensite cracking sites found at lower strain.

Based on the imaging, slicing and registration shown in the methods and results sections above, we give first a visualisation using micrographs of how a damage site evolves across the volume, before considering in more depth the statistical data obtained from the panoramic slice and view at the three strain levels.

Figure 13 illustrates the local information and changing damage type identified in 2D across 8 slices for a damage site extending across five slices that were 0.5 µm apart in the sample strained to 19%. There is no apparent damage in the uppermost two slices (the number of slices in z-direction are indicated in each successive image taking the top left as the reference slice, e.g., ↓x1). As we go deeper, a damage site appears at the depth of 1 µm, in the slice marked with ↓x2. The damage continues to the next slice, marked with ↓x3, and appears as a black pore in both slices showing the characteristics of interface decohesion, i.e. a rounded appearance with clear flow into the ferrite phase. Hence, if we were to analyse each of these slices, we would attribute these damage sites to interface decohesion, where the damage is surrounded partly by the martensite island. However, as we ablate further material, we find that underneath, the two islands appear to have formed a larger block that fractured by a through thickness martensite crack (↓x4, ↓x5). The damage is again identified as interface decohesion in the lower of these slices (↓x5), highlighting the difficulty in classifying evolved damage sites that show signatures of both cracking (parallel fracture surfaces) and flow into the ferrite from the interfaces (at the bottom end of the void imaged in slice ↓x5). In the last slice containing the void,↓x6, it is adjacent to only one martensite island, giving again an interface decohesion site in 2D, before finally disappearing in slice ↓x7.

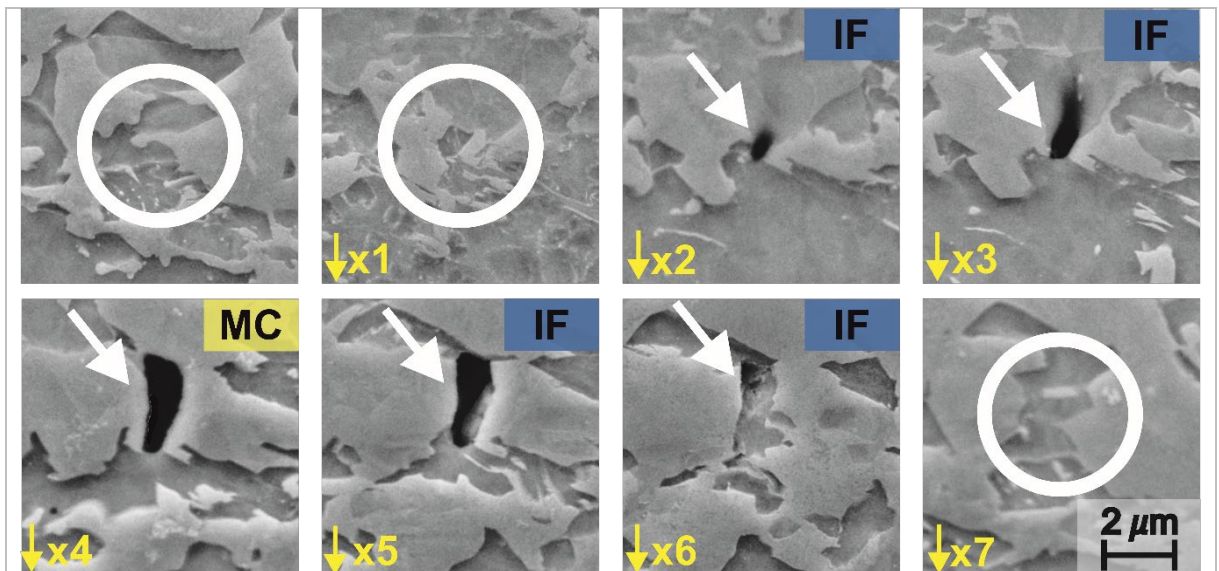

*Figure 13. An example of a damage tracking along the z-axis and illustration of its varying appearance in different slices of the same site. The images belong to the sample with 19% strain and 500 nm slicing thickness. The white circles indicate the damage free surfaces at that specific depth, where the damage is found in the other slices at the same coordinate.*

If we were to look at only individual slices in this case, then the top slice as well as ↓x1 and ↓x7 would suggest that in spite of the closely positioned martensite islands, no damage is induced in the vicinity. In contrast, slices ↓x4 and ↓x5 suggest fracture of a large, blocky martensite island, while slice ↓x6 would imply a plastic decohesion of the ferrite from only one of the two martensite islands. However, the analysis across the three-dimensional evolution of the damage site reveals that the underlying cause of damage is a crack in the central martensite component. We note that, in spite of this, three to four out of the five slices through the damage site suggest flow in the ferrite to dominate. There is therefore a clear difference between what we find is the likely nucleation site and mechanism of damage here and how the site spreads into the volume. This is to be expected and commonly considered as the two stages of damage nucleation and growth. However, a clear challenge arises when these stages cannot be told apart in an analysis. We propose that in 2D view this is not possible and that counting of damage sites alone in individual or independent slices will invariably conflate nucleation and growth.

Considering the new, 3D information available here, we conclude that the transition from martensite cracking to interface decohesion as the dominant mechanism results from continuous nucleation of damage sites by martensite cracking followed by plastic extension of the void into the ferrite, giving rise to interface decohesion morphologies in cross-section. The volume of a given damage site may indeed eventually be dominated by plastic flow/interface decohesion. This is consistent also with the later stage of coalescence across several sites with interacting strain fields, so that a clear tracking of nucleated sites in terms of their number density between straining steps will be difficult at best.

In-situ experiments are the obvious solution to separating stages of an evolving site. However, in-situ deformation inside an SEM suffers from two drawbacks, namely the limitation to 2D, at least during the experiment, and the introduction of severe artefacts due to the zero normal stress at the free surface being imaged [11]. High resolution, non-destructive tomography during straining, using x-rays, may then be of use here. In this case, an interpretation of void evolution may be possible based on void shape evolution alone, even where phase contrast is absent, as long as spatial resolution is sufficiently high. For example, sites with a central component of straight edges are common in the microstructure used here and correspond to damage that originates from martensite cracking and then extends outwards. On the other hand, the interface decohesion sites not associated with a martensite crack may systematically lack this signature of partial parallel faces in tomographic data. The more indirect interpretation of damage site origin in the absence of phase contrast could then be supported by a comparison with the relative fraction of each mechanism identified in a 3D analysis as shown here.

## Damage mechanisms and their interplay across many 3D sites

Next, we consider not only individual, exemplary sites, but many sites - all analysed in 3D and with identified mechanisms tracked for each site individually. In the results shown in Figure 11, containing Sankey diagrams of damage site evolution arranged by strain and void depth, we explore the additional dimension our method offers, namely that we go beyond statistical evaluation of many sites in consecutive slices towards tracking the evolution of each site through the volume, so that transitions in prevalent mechanism appearing in the 2D views are physically connected rather than purely statistical. Owing to the large number of sites contained in our 3D volumes, we can then in turn study the underlying trends revealing the most common 3D morphologies of damage sites and with them the likely interplay of mechanisms giving rise to them.

The distribution of the type of the damage sites along the depth in Figure 11 shows that martensite cracking is not only present as a major mechanism in terms of its fraction at all strains, but we also note that it tends to occur at the centre of damage sites. In almost all cases where the damage site extends to more than two slices, the middle slices are predominantly occupied by martensite cracks, and the relative proportion changes to mostly interface decohesion or notch effect towards the upper or lower boundaries of the damage sites. This becomes more obvious at higher strains with a belly/convex shaped distribution of the yellow area of the Sankey diagrams in Figure 11.

We note that there is an imbalance in this distribution with interface decohesion apparently spreading further up than down into the volume, although the uniaxial stress state would dictate a homogeneous evolution in all directions away from the tensile axis. Two likely causes for this are artefacts from either sample preparation or imaging. The first may be due to a changed stress state too close to a free surface or the introduction of further decohesion during the coarser initial polishing step. As this initial step ablates a considerable amount of material, and cross-sectional imaging through a tensile specimen does not reveal obvious distortion of damage sites towards the free surface, we consider this unlikely in spite of the clear evidence of interference of this polishing step with the first imaged slice (which we discarded in our analysis as a consequence).

Imaging and etching then remain as the likely culprits and we will discuss these further separately. We note here that we do find the greater extension of interface decohesion slices above martensite cracks to be consistent with an imaging artefact. Of course identification of a sequence of events from post-mortem slicing is intrinsically fraught with difficulties. Here, post-mortem slicing cannot alone distinguish for the same void morphology, whether the void nucleated via a brittle crack and spread outwards into the softer ferrite or arose from a stress concentration at the interface and penetrated into the brittle phase. Neither of these variants, however, is consistent with greater extension of interface decohesion identified towards the top, which we attribute to the underlying principle of image formation using secondary electrons. If damage really was to proceed from an original interface site into the martensite and then through the island, one may expect a stronger tendency for sites to divide into two cases where either interface decohesion or martensite dominate the top and bottom, which is not obvious from our data. We therefore conclude that in this majority of cases, void are generated by martensite failure followed by outwards extension through plastic flow.

In fact, martensite damage has previously been identified as not only the major mechanism but as the unique root of damage formation [4].

Our observation of damage sites that are classified as interface decohesion across all slices contradicts this much more general conclusion, at least for the steel investigated here. There are two likely explanations for this: First, we may assume that either all sites are indeed nucleated by martensite cracking under normal stresses but that these sites may be small enough to escape our slicing resolution, resulting in no slice identifying martensite cracking or only identification as notch effect, where also the in-plane resolution is not sufficient to distinguish a fresh but small martensite crack from local plastic flow at protruding martensite arms. However, a second possibility has been raised in the literature [37], resulting not from normal decohesion of martensite islands but rather sliding. Detailed electron channelling contrast imaging (ECCI) have shown the formation of jagged martensite/ferrite interfaces, which are a result of sharp martensite wedges protruding into ferrite. Combining the observations with numerical crystal plasticity modelling has further highlighted the role of substructure boundary sliding of the lath martensite in the initiation of martensite/ferrite interface damage by inducing high plastic deformation localisation in the near-interface ferrite matrix [37].

We note that we do not find a clear sub-class of sites that contain a notch at the centre or a significant bias towards small sites with combination of notch and interface decohesion in adjacent slices, which would be indicative of the first case above. The reason for this may be the large local strain as such a site expands into the surrounding ferrite, giving rise to too few sites where this sequence would remain visible. Based on the present data we therefore cannot infer the importance of decohesion dominated void formation or distinguish such sites from those identified as notch effect elsewhere. The latter form a consistent set of arrangements around or throughout damage sites of all sizes as the branched nature of martensite islands naturally gives rise to these sites throughout the material.

Sliding at pre-existing internal interfaces, as the alternative mechanism leading to perceived decohesion dominance, is not resolved in the current study, as this type of analysis would typically involve more local imaging and analysis of the internal martensite structure [52]. However, the existing data will give the opportunity to filter sites which extend throughout the volume as interface decohesion sites and analyse whether evidence of interface sliding is indeed found. In the current setup of classes and training data used here, this is not possible directly. A thorough analysis of this type of damage nucleation mechanism would have to extend not only the employed classes but ideally also the initial site identification. This is because the early stages of martensite sliding may result in a jagged interface, where the ferrite can still accommodate the local strain, but no cluster of dark pixels will form at this stage, which is used here to identify damage sites. An integrated or additional neural network for finding and classifying sites, such as Yolo [53], or a dedicated edge detection would then have to be used instead of an identification of dark pixel clusters.

As an expansion of such work, statistical damage analysis by imaging can be coupled with crystal plasticity simulation based on phase segmentation (using etching as employed here or EBSD data) as well as high throughput nanoindentation. We would envision the first to yield datasets allowing more reliable reverse identification of parameters for the constitutive material laws and relevant mechanisms that have to be replicated by a given model to yield realistic simulations of damage formation. Nanomechanical mapping may support this further in terms of underlying phase properties and internal gradients developing as a result of dislocation pile-up [54] to explore the relevance of phase contrast [37] in competition with geometric parameters, such as ferrite grain size or martensite island size and substructure boundary length.

## Common artefacts from preparation and imaging

In preparing and analysing the 3D data presented in this work, we considered common artefacts that often go unnoticed or unconsidered in the standard 2D analyses. The two most important ones are (1) the effect of metallographic preparation on the number of damage sites and (2) the effect of imaging conditions on damage measurements in terms of the present mechanisms.

A typical metallographic preparation consists of grinding of the surface with a sequence of sanding papers of various coarseness and subsequent polishing steps, usually similar to what was applied here. Grinding times are rarely given and are in any case likely of limited use as long as applied force, rotational/radial speed of the polishing wheel and lubrication conditions are not fixed. We found here, where we followed a standard initial preparation step with successive fine polishing steps, that the first slice following grinding results in more damage compared to all subsequent slices. Mechanical damage in terms of the introduction of dislocations during metallographic preparation is well known, for example in nanomechanical testing, where the prominent pop-in associated with dislocation nucleation disappears upon polishing for material with originally low dislocation density [55] . Beyond such an introduction of a thin layer of increased dislocation density, the introduction of cold working and accordingly micro grooves on the surface may affect several surface properties at once, such as the roughness of the surface, dislocation density, work hardening state and residual stress [56].

While it is beyond the scope of this work to identify the exact fraction of damage sites induced by this procedure, the increase of damage sites on the initial surface layer compared to all subsequent slices is such that it must be considered where a comparison of damage density is made between different laboratories or even different people following a metallographic routine. It appears likely that a change of routine, e.g., reduction in final grinding grit size or applied force or increase in polishing times, would have a greater impact on damage density than the parameters often considered of interest, such as a difference in applied strain or microstructure. With reference to Figure 8 we note that we achieve approximately the same number of damage sites across the same measured area for the first slice after 14% strain and the lower slices at 25% strain. Owing to the availability of several slices and a carefully controlled slicing procedure in our case, we can identify this artefact and judge that it does not appear to extend past the first or at most second slice. Below this point, consistent damage densities are found. However, where such a depth profile is not available, constant preparation routines are essential for relative comparisons of different samples and deep and careful polishing is imperative, if representative absolute damage measurements are important. In the absence of controlling these preparation parameters, a strong bias or experimental scatter exceeding the effect of controlled experimental parameters must be expected.

As part of our 3D analysis of the damage mechanisms identified in the 2D slices, we further noticed that deeper damage sites show a tendency to develop more interface decohesion towards the top (Figure 11). This is in fact consistent with two artefacts stemming from imaging inside an SEM. First, we use secondary electron contrast to resolve the topographic contrast introduced as a result of etching and damage as the main signal. Secondary electron yield is greatly affected by edges, such that sites with a steep inclination on at least one side tend to be more visible owing to the greater variation in secondary electron escape. Where a damage site protrudes into the volume with a rounded shape at the bottom, the last slice may only contain a shallow basin that remains from the original site. This will be the case where plasticity in the ferrite locally dominates, even where it is driven from an adjacent martensite crack. As such a shallow basin leads to reduced contrast compared to, for example, a martensite crack with sharp edges of the same normal depth, the applied imaging method is prone to overlooking these sites. As a result, the apparent depth of interface decohesion sites in particular is reduced. In contrast, the sites positioned towards the top are artificially increased. Where we image such a damage site that has not been opened by either polishing or etching but is positioned directly underneath the current surface, the interaction volume of the electron beam may already penetrate the void. In this case, a lower secondary electron yield will be the result. This is detected as dark pixels and hence particularly large damage sites may give rise to the appearance of an extended but more likely smeared appearance towards the top, resulting in the identification as a decohesion or notch site depending on lateral extension. The imaging method therefore contributes to the observed imbalance of plastic extension of martensite cracks towards the top and bottom in both ways. In addition, preferential etching and polishing may further contribute by opening damage sites more strongly from the top than expanding them towards the bottom.

At the very least qualitatively, it is well known how these artefacts arise and how they can be controlled better. Comparing different detector signals and/or changing acceleration voltage may give important information on imaging artefacts. Here, we chose imaging with secondary electrons at 20 kV to achieve fast and reliable imaging conditions for automatic image acquisition with clear phase contrast at reasonable resolution. As in metallographic preparation, it is imperative to keep these conditions constant across datasets and to carefully choose imaging conditions as a good compromise between the most vital signals, their strength and noise.

# Conclusions

In this work, we analyse the three-dimensional morphology, distribution and depth of damage sites in dual-phase steel after straining to 14%, 19%, and 25% elongation. We were able to track the behaviour and evolution of damage for thousands of sites in three dimensions using high-resolution panoramic electron micrographs with an area of 850 µm * 850 µm across ten slices 0.25 - 0.5 µm apart into the depth of the material, as well as careful image registration and automated damage mechanism analysis by convolutional neural networks.

By analysing this large number of sites and tracking each site's evolution individually, we avoid the typical selection bias in imaging small and manually selected areas. As a result, this method provides and confirms insights into the formation of damage voids during deformation.

1) We demonstrate that the perceived type and evolution of damage changes strongly along each site's depth and that the apparent structure of a damage site may be misleading in a single two-dimensional view.
2) We identify martensite cracking as the dominant mechanism leading to damage formation, while ferrite plasticity leads to growth of these sites. In 2D analyses, this may be perceived as interface decohesion dominating damage, particularly at larger strains.
3) Our 3D analysis further revealed that martensite cracking is unlikely to be the sole origin of damage nucleation, as the initial stages of martensite sliding are much more difficult to detect and track. This is supported by a fraction of sites dominated by ferrite plasticity without visible evidence of martensite cracking within our spatial resolution.
4) In the investigated microstructure, consisting of strongly branched martensite, damage site depth is randomly distributed without a significant internal length scale down to the order of a few hundred nm. As strain increases, the number of sites increases, but no preference in site depth emerges.
5) In terms of metallographic sample preparation, we find that grinding and insufficient polishing may induce artefacts in the number of damage sites that easily outweigh differences in the applied strain or investigated microstructure.

# Acknowledgement

The authors gratefully acknowledge funding by the Deutsche Forschungsgemeinschaft through Collaborative Research Center TRR 188 (project ID 278868966) and Collaborative Research Center CRC 1394 - Structural and Chemical Atomic Complexity: From Defect Phase Diagrams to Material Properties (project ID 409476157). This work is also funded by the Federal Ministry of Education and Research (BMBF) and the state of North Rhine-Westphalia as part of the NHR Program. Calculations were performed with computing resources granted by RWTH Aachen University under project rwth0535. We further thank Dr. Ing. Carl Kusche for providing the SEM image of the cross-section sample.

# Code and Data Availability

The codes and some data will be publicly accessible within a Zenodo repository in the final version once the paper in accepted and we are happy to share the large image dataset on request to be able to offer support also in its use and analysis. They are now stored in private repository in the first author's github.

# Supplemental Material

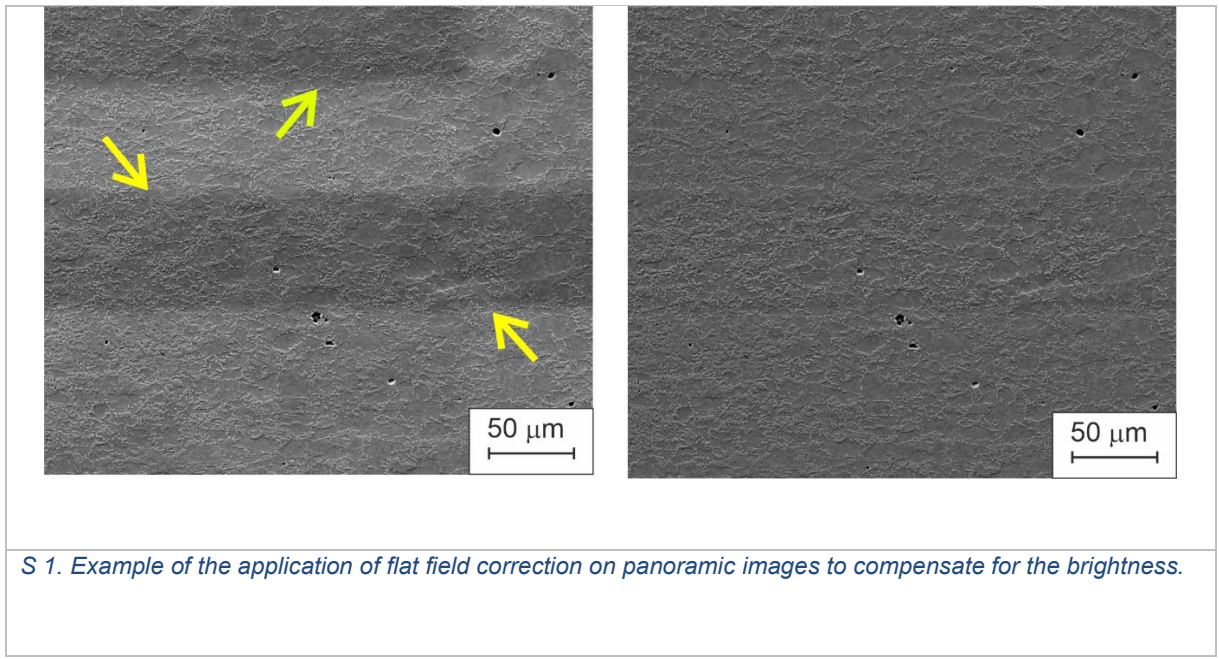

*S 1. Example of the application of flat field correction on panoramic images to compensate for the brightness.*

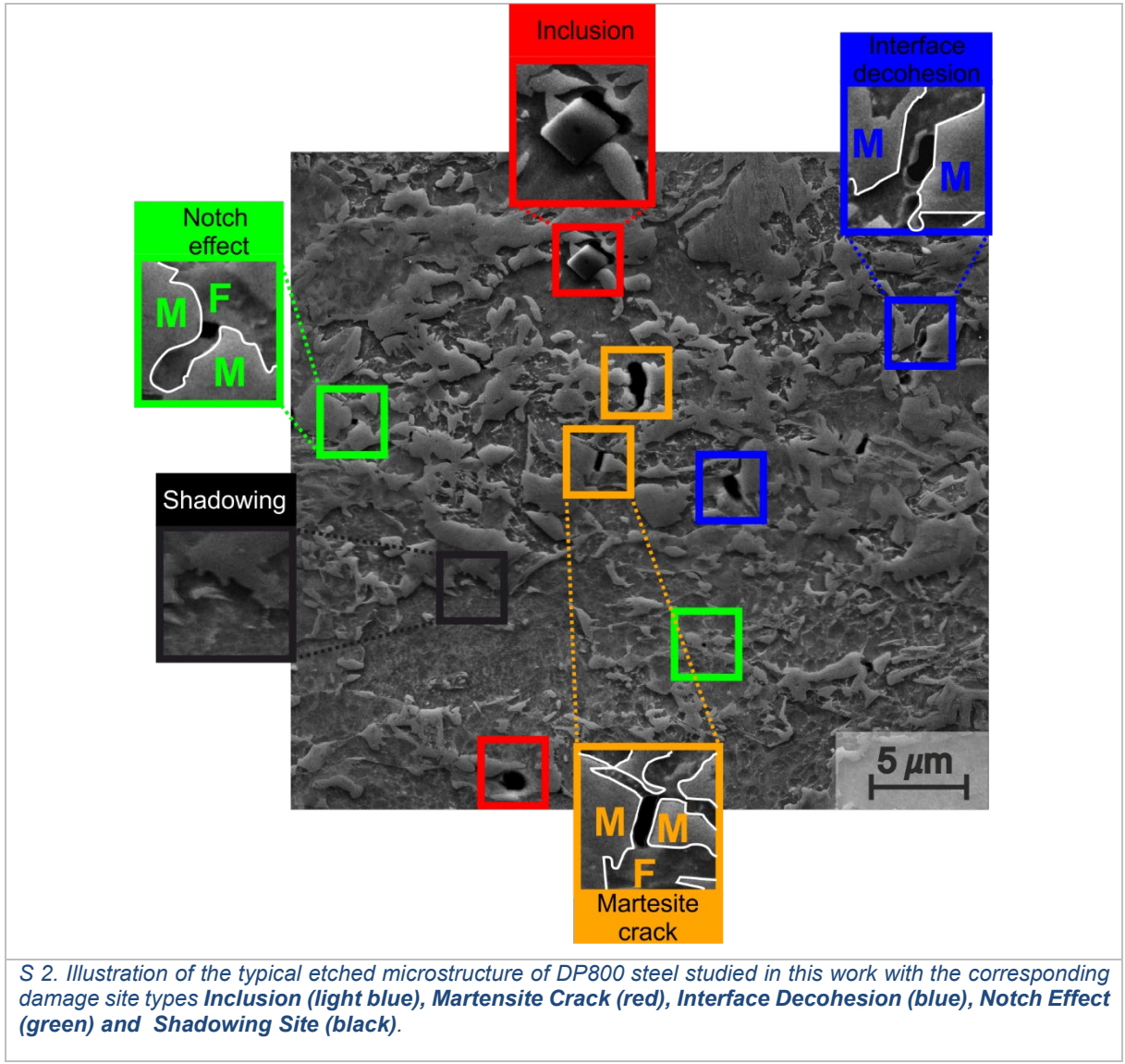

*S 2. Illustration of the typical etched microstructure of DP800 steel studied in this work with the corresponding damage site types **Inclusion (light blue), Martensite Crack (red), Interface Decohesion (blue), Notch Effect (green) and Shadowing Site (black)**.*

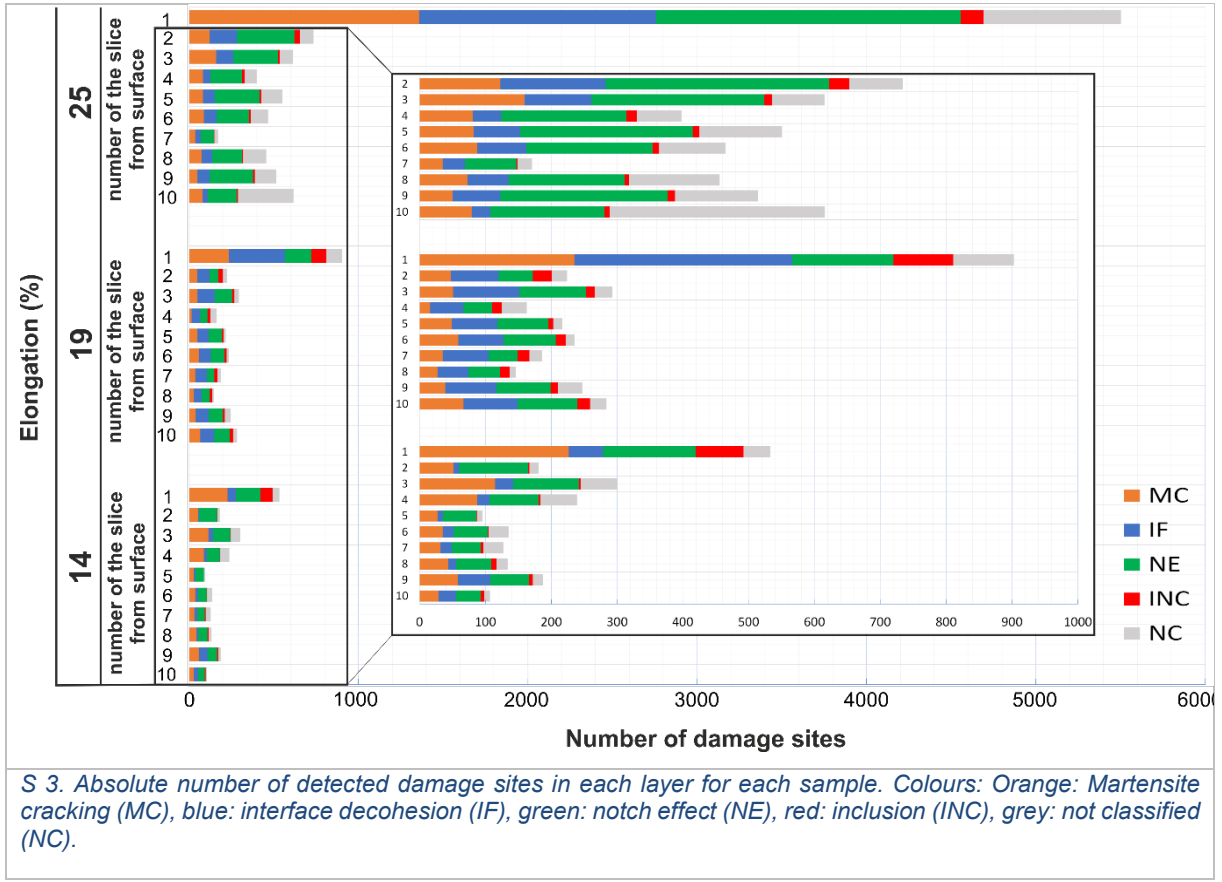

*S 3. Absolute number of detected damage sites in each layer for each sample. Colours: Orange: Martensite cracking (MC), blue: interface decohesion (IF), green: notch effect (NE), red: inclusion (INC), grey: not classified (NC).*

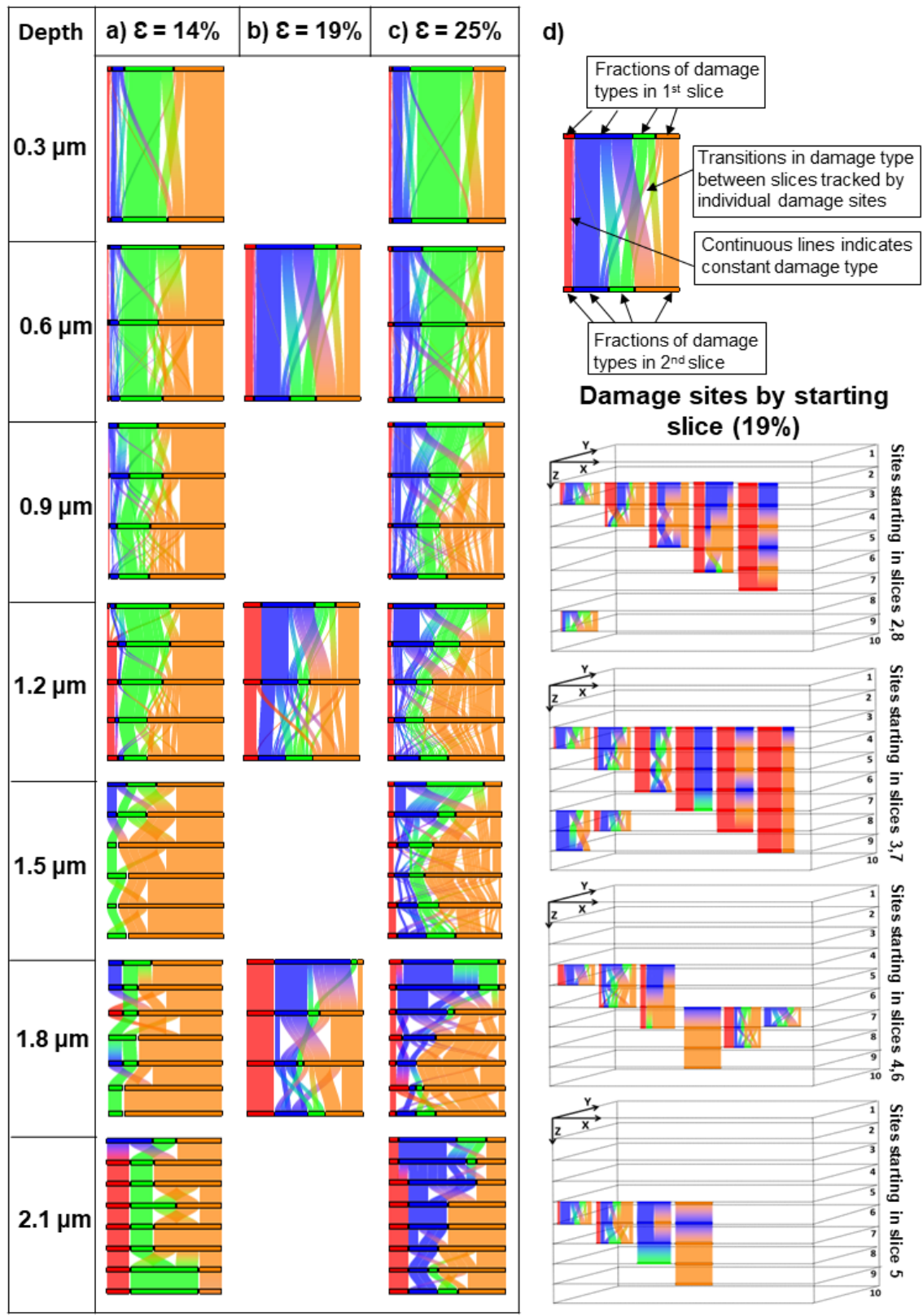

*S 4. Sankey diagrams by sample elongation and damage site depth tracking the 2D classification of damage sites in 3D. Colours identify damage type; horizontal bar width indicates type fraction at the given site depth and vertical connections trace damage type evolution along the depth for all damage sites individually. Colours (from left to right): Red: inclusion, blue: interface decohesion, green: notch effect, orange: martensite cracking.*

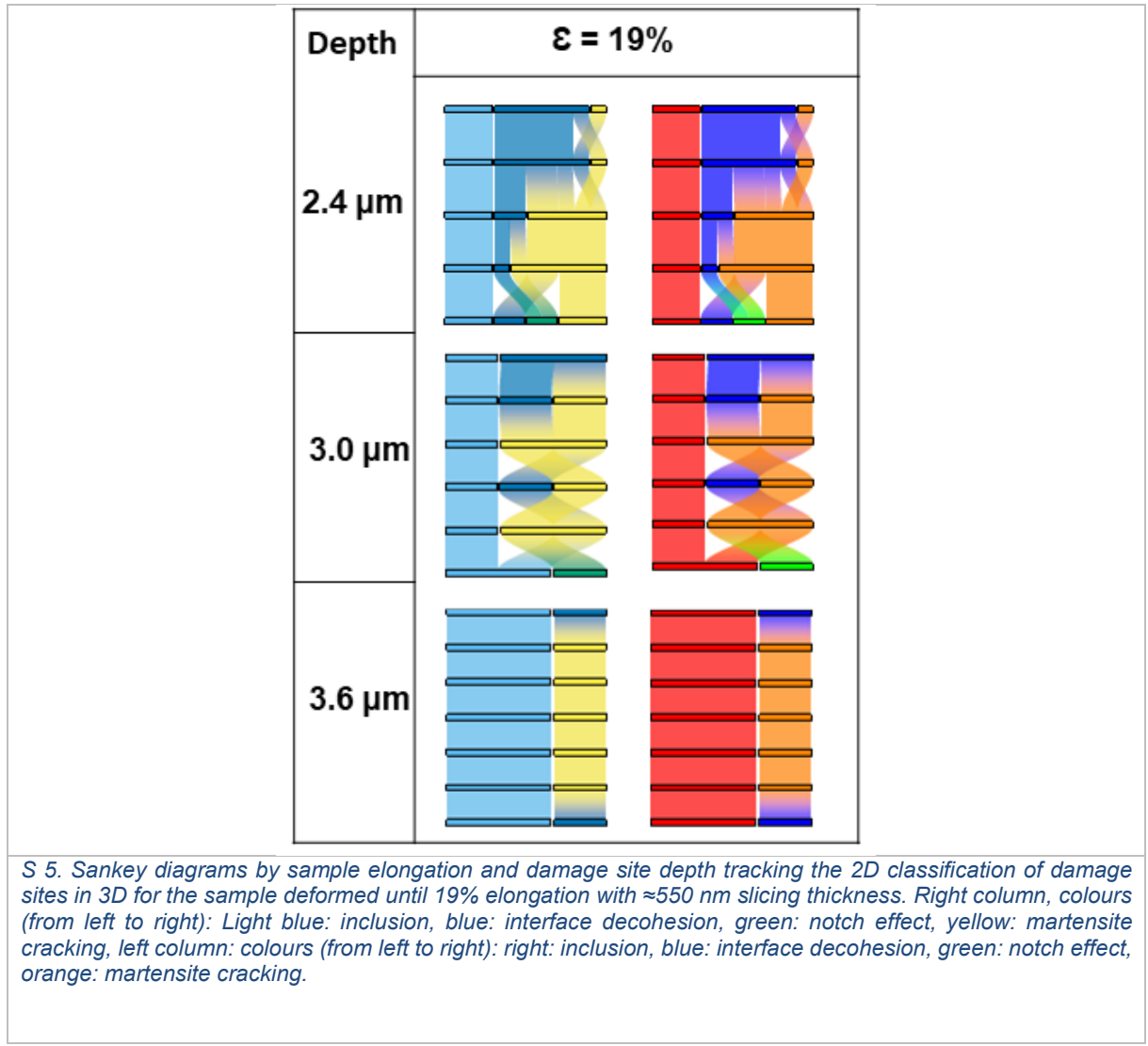

*S 5. Sankey diagrams by sample elongation and damage site depth tracking the 2D classification of damage sites in 3D for the sample deformed until 19% elongation with ≈550 nm slicing thickness. Right column, colours (from left to right): Light blue: inclusion, blue: interface decohesion, green: notch effect, yellow: martensite cracking, left column: colours (from left to right): right: inclusion, blue: interface decohesion, green: notch effect, orange: martensite cracking.*